\documentclass[journal]{IEEEtran}

\usepackage{mathrsfs}
\usepackage[noadjust]{cite}
\usepackage{graphicx,color,overpic,psfrag}
\usepackage{float}
\usepackage{stfloats}
\usepackage{amsmath, amssymb}
\usepackage{cleveref}
\usepackage{latexsym}
\usepackage{bm}
\usepackage{amssymb}
\usepackage{cases}
\usepackage{array}
\usepackage{fancyhdr}
\usepackage{setspace}

\ifCLASSOPTIONcompsoc
\usepackage[caption=false,font=normalsize,labelfont=sf,textfont=sf]{subfig}
\else
\usepackage[caption=false,font=footnotesize]{subfig}
\fi

\definecolor{myblue}{rgb}{.91,.95,.99}
\usepackage{multirow}
\usepackage{url}
\usepackage{algorithm}
\usepackage{algorithmic}

\usepackage{blkarray}
\usepackage{booktabs}

\usepackage{multirow}
\usepackage{dsfont}
\usepackage{tabularx}
\usepackage[table]{xcolor}
\usepackage{amsfonts}

\usepackage{letltxmacro}

\graphicspath{{figure/}}

\newcommand{\ra}[1]{\renewcommand{\arraystretch}{#1}}
\newcolumntype{L}{>{\hspace*{-\tabcolsep}}l}
\newcolumntype{R}{c<{\hspace*{-\tabcolsep}}}
\definecolor{lightblue}{rgb}{0.93,0.95,1.0}

\newcommand{\figref}[1]{Fig. \ref{#1}}
\newcommand{\tabref}[1]{Table \ref{#1}}

\newcommand{\appref}[1]{Appendix \ref{#1}}

\newcommand{\diag}[1]{\mathsf{diag}\left\{#1\right\}}
\newcommand{\blkdiag}[1]{\mathsf{blkdiag}\left\{#1\right\}}

\newcommand{\abs}[1]{\left|#1\right|}

\newcommand{\thetabs}[2]{{\dnnot{\theta}{bs}}}

\newcommand{\cB}{\mathcal{B}}

\newcommand{\cP}{\mathcal{P}}
\newcommand{\cQ}{\mathcal{Q}}

\newcommand{\cS}{\mathcal{S}}

\newcommand{\cU}{\mathcal{U}}

\newcommand{\ba}{\mathbf{a}}
\newcommand{\bb}{\mathbf{b}}

\newcommand{\bh}{\mathbf{h}}

\newcommand{\bp}{\mathbf{p}}

\newcommand{\bs}{\mathbf{s}}

\newcommand{\bu}{\mathbf{u}}
\newcommand{\bv}{\mathbf{v}}
\newcommand{\bw}{\mathbf{w}}
\newcommand{\bx}{\mathbf{x}}
\newcommand{\by}{\mathbf{y}}
\newcommand{\bz}{\mathbf{z}}

\newcommand{\bA}{\mathbf{A}}
\newcommand{\bB}{\mathbf{B}}
\newcommand{\bC}{\mathbf{C}}
\newcommand{\bD}{\mathbf{D}}

\newcommand{\bG}{\mathbf{G}}
\newcommand{\bH}{\mathbf{H}}
\newcommand{\bI}{\mathbf{I}}

\newcommand{\bQ}{\mathbf{Q}}
\newcommand{\bR}{\mathbf{R}}

\newcommand{\bT}{\mathbf{T}}
\newcommand{\bU}{\mathbf{U}}

\newcommand{\bW}{\mathbf{W}}
\newcommand{\bX}{\mathbf{X}}
\newcommand{\bY}{\mathbf{Y}}
\newcommand{\bZ}{\mathbf{Z}}

\newcommand{\bbC}{\mathbb{C}}

\newcommand{\sinr}{\mathtt{SINR}}

\newcommand{\dnnot}[2]{#1_{\mathrm{#2}}}

\allowdisplaybreaks

\begin{document}

\title{\huge Beam Squint-Aware Integrated Sensing and Communications for Hybrid Massive MIMO LEO Satellite Systems}

\author{
Li~You,
Xiaoyu~Qiang,
Christos~G.~Tsinos,
Fan~Liu,
Wenjin~Wang,
Xiqi~Gao,
and~Bj\"{o}rn~Ottersten

\thanks{Copyright (c) 2015 IEEE. Personal use of this material is permitted. However, permission to use this material for any other purposes must be obtained from the IEEE by sending a request to pubs-permissions@ieee.org.}
\thanks{Part of this work was submitted to ICC'2022 Workshop \cite{qiang2022joint}.}
\thanks{
Li You, Xiaoyu Qiang, Wenjin Wang, and Xiqi Gao are with the National Mobile Communications Research Laboratory, Southeast University, Nanjing 210096, China, and also with the Purple
Mountain Laboratories, Nanjing 211100, China (e-mail: lyou@seu.edu.cn; xyqiang@seu.edu.cn; wangwj@seu.edu.cn; xqgao@seu.edu.cn).
}
\thanks{
Christos G. Tsinos and Bj\"{o}rn Ottersten are with the University of Luxembourg, Luxembourg City 2721, Luxembourg (e-mail: chtsinos@gmail.com; bjorn.ottersten@uni.lu).
}
\thanks{Fan Liu is with the Department of Electronic and Electrical Engineering, Southern University of Science and Technology, Shenzhen 518055, China (e-mail: liuf6@sustech.edu.cn).
}
}
\maketitle

\begin{abstract}
  The space-air-ground-sea integrated network (SAGSIN) plays an important role in offering global coverage.
  To improve the efficient utilization of spectral and hardware resources in the SAGSIN, integrated sensing and communications (ISAC) has drawn extensive attention.
  Most existing ISAC works focus on terrestrial networks and can not be straightforwardly applied in satellite systems due to the significantly different electromagnetic wave propagation properties.
  In this work, we investigate the application of ISAC in massive multiple-input multiple-output (MIMO) low earth orbit (LEO) satellite systems.
  We first characterize the statistical wave propagation properties by considering beam squint effects.
  Based on this analysis, we propose a beam squint-aware ISAC technique for hybrid analog/digital massive MIMO LEO satellite systems exploiting statistical channel state information.
  Simulation results demonstrate that the proposed scheme can operate both the wireless communications and the target sensing simultaneously with satisfactory performance, and the beam-squint effects can be efficiently mitigated with the proposed method in typical LEO satellite systems.
\end{abstract}


\section{Introduction}\label{sec:net_intro}

The increasing demand for services in the sparsely populated or the un-deployed areas, i.e., marine and aeronautical regions, motivates the study of the space-air-ground-sea integrated network (SAGSIN) \cite{liao2021terahertz,liu2021leo,you2021towards,xiao2020uav}.
The SAGSIN involves not only terrestrial networks but also spaceborne, airborne, and marine parts, thus offering global coverage.
The spaceborne part of the SAGSIN consists of numerous satellites deployed at different altitudes. Geostationary satellites require large investments, high launch costs, and suffer from high propagation delay, which leads the interest for the non-geostationary counterparts, including the low earth orbit (LEO) satellites \cite{al2021broadband}.
The 500--2000 km orbit altitudes of the LEO satellites are relatively lower than the medium earth orbit (MEO) or high elliptical orbit (HEO) ones, leading to lower latency and higher data rates for wireless communications \cite{qu2017leo}.

With the great development of the wireless communication industry, spectrum resources tend to be increasingly limited and thus valuable. 
To improve the utilization efficiency of the precious spectrum resources, integrated sensing and communications (ISAC) is proposed as a way to achieve frequency reuse between the two functional modules, i.e., wireless communications and sensing \cite{liu2020joint,zhang2018multibeam}.\footnote{In the literature, ISAC is also referred to as joint communications-sensing (JCS), joint radar communications (JRC), joint communications and radar sensing (JCAS), dual-functional radar communications (DFRC), etc.}
In the ISAC system, communications and sensing can be performed simultaneously in one hardware platform, enabling the decongestion of the radio frequency (RF) environment \cite{liu2020joint}.

The existing ISAC works mainly focus on terrestrial networks and have explored many promising applications, e.g., massive multiple-input multiple-output (MIMO) \cite{liu2020joint}, to improve the performance of both modules.
However, despite the promising performance gains due to the massive MIMO technology, the massive number of antennas might lead to frequency-dependent array responses and cause  severe beam squint effects, which can be mitigated with proper design \cite{cheng2021hybrid,xu2021wideband}.
Besides, in the existing terrestrial ISAC systems, hybrid transceivers integrated with a sub-arrayed MIMO radar are usually combined with the massive MIMO technology to reduce the number of RF chains \cite{liu2019hybrid,liu2020joint,kaushik2021hardware}.

In this work, we propose to operate the ISAC in the LEO satellite systems, which has great potential in providing wide coverage for wireless communications and sensing, and presents great compatibility with the SAGSIN.
It is worth noting that the previous ISAC works for terrestrial systems can not be directly adopted in the considered LEO satellite systems due to the significantly different wave propagation properties.
In particular, there exist two major differences to be highlighted, i.e., the inevitably high propagation delay and large Doppler shifts due to the long distances between the LEO satellites and the user terminals (UTs)/targets as well as their mobility \cite{you2020massive,cherniakov2002air}.
Moreover, the consideration of a wide-band massive MIMO LEO satellite ISAC system involves the adoption of a large array and wide bandwidth, leading to high-dimensional and rapid-varying channel.
Due to the above reasons, the accurate instantaneous channel state information (iCSI) at the transmitter is practically difficult to be estimated in the considered satellite ISAC system.
Thus, we propose to design the considered satellite ISAC system based on the statistical CSI (sCSI), which varies significantly less on small time scales.
Note that in the existing LEO satellite communication (SATCOM) systems, the sCSI has already been investigated due to the slow-varying property \cite{you2020massive}.

Motivated by the above considerations, we propose a beam squint-aware hybrid analog/digital transmitter for ISAC in massive MIMO LEO satellite systems based on sCSI. To the authors' best knowledge, this is the first work that investigates the adoption of ISAC in satellite systems.
In particular, the contributions of our work are summarized as follows:
\begin{itemize}
  \item We characterize the statistical wave propagation properties by considering beam squint effects, that appear in the proposed massive MIMO LEO satellite ISAC scenario. Besides, we identify the relationship between the typical system parameters of the LEO satellite ISAC system (i.e., the system bandwidth, the carrier frequency, the aperture of the antenna array) and the beam squint effects.
  \item We design the transmitter of the proposed LEO satellite ISAC system, to simultaneously perform communications and sensing.
       In particular, a weighting coefficient is introduced to adjust the weight between these two functional modules and enable a trade-off between their performance, which is measured by the energy efficiency (EE) of the communication part and the sensing beampattern, respectively.
  \item We develop an efficient algorithmic approach with the utilization of sCSI knowledge for hybrid precoding in the LEO satellite ISAC system, to mitigate the beam squint effects and enhance the communication EE as well as the sensing beampattern matching performance, respectively.
\end{itemize}
\subsection{Related Works}
\emph{LEO SATCOM} -- So far, for LEO SATCOM systems, the signal propagation properties and massive MIMO uplink/downlink transmission based on the sCSI knowledge have been studied in \cite{you2020massive,li2020downlink}.
In addition, in LEO satellite systems, the hardware restriction and power supplement mechanism impose the significance of considering the EE metric, which trades off between the downlink data rate and the power consumption at the transmitter \cite{fraire2020battery}.
To that end, in \cite{gao2021sum,you2022hybrid}, the authors investigated the adoption of the hybrid precoding scheme into the LEO SATCOM systems to improve the sum rate or the EE performance at a reduced number of RF chains.

\emph{Massive MIMO Radar} -- Sub-arrayed MIMO radars, which combine the advantages of the phase-arrayed and MIMO radars, have been recently attracted great interest in terrestrial systems \cite{liu2019hybrid,liu2020joint,kaushik2021hardware}.
In particular, in the phase-arrayed radar system, an identical signal is transmitted from all the antennas, and thus, only one RF chain is required, resulting in high array gain \cite{wilcox2011mimo} and low hardware complexity as well as power consumption \cite{kaushik2021hardware}.
However, the employed array with massive antennas potentially offers a more efficient utilization. Thus, a MIMO radar system is proposed where independent signals are transmitted from different antennas.
Then, the number of required RF chains is equal to that of the antennas.
Therefore, the sub-arrayed MIMO radar is proposed with the view to find a compromise between these two architectures via dividing the antenna array into several non-overlapping subarrays \cite{hassanien2010phased}.

\emph{Terrestrial ISAC} -- Many previous works have investigated ISAC design in terrestrial networks.
In  \cite{liu2020joint}, the authors overviewed the existing application scenes as well as the technology advances, and proposed a DFRC system.
Furthermore, in \cite{liu2018toward,liu2019hybrid,kaushik2021hardware}, the authors focused on the design of a hybrid beamformer with different MIMO radar techniques for the DFRC system.
Recently, some promising operations have been investigated for terrestrial ISAC systems.
In particular, orthogonal frequency division multiplexing (OFDM) signals are adopted for communications to mitigate the inter-symbol interference, which can also be employed for target sensing \cite{ahmed2019ofdm,tian2021transmit}.
Besides, the multibeam technology has been proposed to satisfy various requirements for beamwidth and power levels for the two functional modules \cite{zhang2018multibeam}.
Moreover, the MIMO technology has been adopted to offer great spatial degrees of freedom and provides compensation for path loss.
Thus, it can significantly improve the spectral efficiency (SE) and EE for the communication module \cite{liu2020joint}.
In addition, for the radar module, the massive MIMO technology has potential benefits in improving the resolution of target sensing as well as enhancing the robustness in the case of the unknown disturbance \cite{liu2020joint}.

\emph{Beam Squint} -- In the terrestrial wide-band massive MIMO communications and/or radar systems, the large array poses a challenge and may lead to significant performance degradation.
Specifically, the propagation delay across the array tends to be non-negligible for the large array.
Therefore, the array response actually varies across the subcarriers, due to not only the multipath channel fading, but also the propagation delay across the array, which can be interpreted as a disturbance imposed on the beam direction \cite{dovelos2021channel}.
This phenomenon is termed beam squint, which is also known as the spatial wide-band effect.
In fact, the beam squint effect has been a case of study since the early radar systems and have been extensively studied since then \cite{yoon2006tops,fan2019mimo}.
Besides, the effects of beam squint have been intensively investigated in the terrestrial communication systems \cite{wang2018spatial,chen2020hybrid,dovelos2021channel}.
To mitigate beam squint effects, the design of the hybrid precoder and combiner has been studied in the wide-band MIMO communication systems in \cite{sohrabi2017hybrid,gonzalez2019hybrid,chen2020hybrid,dovelos2021channel}.
Moreover, the beam squint-aware DFRC systems have recently attracted wide attention in the design of beamforming \cite{cheng2021hybrid,xu2021wideband}, based on the iCSI, which is usually difficult to obtain at the transmitter of the LEO satellite systems, as discussed above.

\subsection{Organization}
The organization of the remaining of this paper is summarized as follows.
The LEO satellite ISAC system model with respect to the communication and sensing modules is presented in Section \ref{sec:sys_mod} and an optimization problem is formulated to make a trade-off between these two modules.
Section \ref{sec:bs_fdp} focuses on the corresponding equivalent fully digital problem with the consideration of the beam squint effects.
Section \ref{sec:JCS_hp} develops algorithms to design the hybrid analog/digital precoders for the ISAC system, implemented with the fully and partially connected structures, respectively.
The simulation results are discussed in Section \ref{sec:sim} and Section \ref{sec:conc} makes a brief conclusion of the paper.

\subsection{Notations}
The upper and lower case boldface letters represent the matrices and column vectors, respectively.
The definition of the imaginary unit is given by $\jmath=\sqrt{-1}$.
The representation of unitary space with $m\times n$-dimension is shown as $\mathbb{C}^{m\times n}$.
The right hand side of $\triangleq$ denotes the definition of the left hand side.
The operator of the Kronecker product of two matrices is represented by $\otimes$.
The symbols $\exp\{\cdot\}$ and $\log\{\}$ stand for the exponential and logarithmic operations, respectively.
The $N\times N$ identity matrix is denoted by $\bI_{N}$.
The representations of operating the transpose, conjugate, and Hermitian conjugate are presented as $(\cdot)^T$, $(\cdot)^\ast$, and $(\cdot)^H$, respectively.
The operations $|x|$, $\angle x$, and $\lceil x \rceil$ stand for taking the amplitude, the angel, and the ceiling value of the inside number $x$, respectively.
The symbol $\mathcal{CN}(\mathbf{0},\sigma^2)$ denotes the circular symmetric complex-valued zero-mean Gaussian distributed scalar with variance $\sigma^2$ and the uniform distribution in the interval $[a,b)$ is denoted as $\cU(a,b)$.
We represent the expectation and the trace operators as $\mathbb{E}\{\cdot\}$ and ${\rm Tr}\left\{\cdot\right\}$.
The block diagonal matrix is denoted as $\bA=\blkdiag{\bX_1,\ldots,\bX_N}$ where the elements $\bX_1,\ldots,\bX_N$ are block matrices on the principal diagonal of matrix $\bA$.
The denotations of $\ell_2$-norm and Frobenius-norm are given by $||\cdot||_2$ and $||\cdot||_F$, respectively.
The symbol $\left[\bA\right]_{i,j}$ stands for the $(i,j)$th element in matrix $\bA$.
We adopt $\varnothing$ to express the empty set.


%
%
%
\section{System Model and Problem Formulation}\label{sec:sys_mod}
\begin{figure}[!t]
		\centering
		\includegraphics[width=0.45\textwidth]{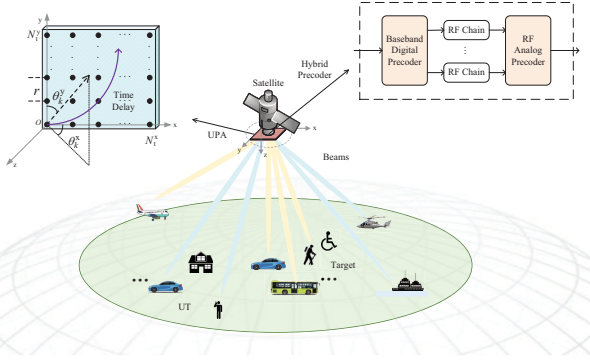}
		\caption{Joint communications and sensing for satellite systems.}
        \label{jsatradsym}
	\end{figure}

Consider an LEO satellite ISAC system providing communication services for $K$ single-antenna UTs as well as detecting targets, as depicted in \figref{jsatradsym}.
The carrier frequency and wavelength of the system are denoted as $f_c$ and $\lambda_c$, where $f_c=c/\lambda_c$ and $c$ denotes the speed of light.
The system bandwidth is represented by $B_{\rm w}$ and the signal duration is estimated as $T_{\rm s}= 1/B_{\rm w}$.
The satellite side is equipped with a uniform planar array (UPA) with $N_\mathrm{t}=N_\mathrm{t}^{\mathrm{x}}\times N_\mathrm{t}^{\mathrm{y}}$ antennas,
where $N_\mathrm{t}^{\mathrm{x}}$ and $N_\mathrm{t}^{\mathrm{y}}$ are the numbers of antennas on the x- and y-axes, respectively.
Note that due to the adopted massive MIMO technology, the number of the antennas $N_{\rm t}$ can be large.
Besides, the antenna separation on both the x- and y-axes
is set to be equal, i.e., $r_{\rm x}=r_{\rm y}=r$.

Due to the frequency selectivity in the considered wide-band massive MIMO LEO system, the OFDM scheme is employed to mitigate the inter-symbol interference \cite{you2020massive,tian2021transmit}.
In particular, a total of $M$ subcarriers are employed over the signal bandwidth $B_{\rm w}$. Then, the $m$th subcarrier frequency is denoted by
    \begin{align}\label{eq:ofdmf}
    f_m=\left(m-\frac{M+1}{2}\right)\Delta_B,\ m=1,2,\cdots,M,
    \end{align}
where $\Delta_B= B_{\rm w}/M$ is the subcarrier separation.

\subsection{Channel Model with Beam Squint Effects in the Communication Module}\label{subsec:chmod}
Generally, the downlink channel of the satellite ISAC systems is characterized by the multipath propagations.
In addition, due to the greatly higher altitude of the satellites compared with the surrounding scatterers of the UTs in the terrestrial side, the angle-of-departure (AoD) of each propagation path can be assumed to be the same \cite{you2020massive}.
Then, the overall delay between the $k$th UT and the $(n_{\rm x},n_{\rm y})$th element of the antenna array through the $l$th path is expressed as
\begin{align}\label{eq:overalldelay}
\tau_{k,l,n_{\rm x},n_{\rm y}}=\tau_{k,l}+\tau_{n_{\rm x},n_{\rm y}}\left(\bm{\vartheta}_k\right),
\end{align}
where $\tau_{k,l}$ is the propagation delay of UT $k$ over the $l$th path. The second component at the right hand side of \eqref{eq:overalldelay} denotes the time delay for the $k$th UT from the $(1,1)$th to the $(n_{\rm x},n_{\rm y})$th element of the antenna array, given by
\begin{align}\label{eq:tmdl}
\tau_{n_{\rm x},n_{\rm y}}\left(\bm{\vartheta}_k\right)\triangleq\frac{r\left(\left(n_{\rm x}-1\right)\vartheta_k^{\rm x}+\left(n_{\rm y}-1\right)\vartheta_k^{\rm y}\right)}{c},
\end{align}
where $n_{\rm x}\in \left\{1,2,\cdots, N_{\rm t}^{\rm x}\right\}$ and $n_{\rm y}\in \left\{1,2,\cdots, N_{\rm t}^{\rm y}\right\}$.
Then, the space angle pair $\bm{\vartheta}_k=\left(\vartheta_k^{\mathrm{x}},\vartheta_k^{\mathrm{y}}\right)$ in \eqref{eq:overalldelay}
can be characterized by the AoD pair $\left(\theta_k^{\mathrm{x}},\theta_k^{\mathrm{y}}\right)$, where $\vartheta_k^{\mathrm{x}}=\sin{\theta_k^{\mathrm{y}}}\cos{\theta_k^{\mathrm{x}}}$ and $\vartheta_k^{\mathrm{y}}=\cos{\theta_k^{\mathrm{y}}}$ \cite{you2020massive}.

Subsequently, utilizing the ray-tracing approach \cite{you2020massive}, the received baseband signal (in absence of noise) from the $(n_{\rm x},n_{\rm y})$th antenna element to the $k$th UT is given by
\begin{align}\label{eq:bbrs}
r_{k,n_{\rm x},n_{\rm y}}(t)=\sum_{l=1}^{L_k}\alpha_{k,l}x_k\left(t-\tau_{k,l,n_{\rm x},n_{\rm y}}\right)\exp\left\{\jmath2\pi t\nu_{k,l}\right\}\notag\\
\exp\left\{-\jmath2\pi f_c\tau_{n_{\rm x},n_{\rm y}}\left(\bm{\vartheta}_k\right)\right\},
\end{align}
where $x_k$ is the transmitted signal for the $k$th UT, $L_k$ denotes the number of channel paths of the $k$th UT, $\alpha_{k,l}$ is the corresponding channel gain. The Doppler shift $\nu_{k,l}$ can be mainly represented by the sum of Doppler shifts coming from the mobility of the satellites and UTs, i.e., $\nu_{k,l}=\nu_{k,l}^{\rm sat}+\nu_{k,l}^{\rm ut}$ \cite{you2020massive}. Note that $\nu_{k,l}^{\rm sat}$ is nearly the same for each channel path $l$ of the $k$th UT, i.e., $\nu_{k,l}^{\rm sat}=\nu_k^{\rm sat}$, due to the high altitude of the satellites \cite{you2020massive}.

Based on Eq. \eqref{eq:bbrs}, the space-time variant response of the downlink channel between the $k$th UT and the $(n_{\rm x},n_{\rm y})$th antenna element of the LEO satellite can be modeled as
\begin{align}\label{eq:httau}
h_{k,n_{\rm x},n_{\rm y}}(t,\tau)=\sum_{l=1}^{L_k}\alpha_{k,l}\delta\left(\tau-\tau_{k,l,n_{\rm x},n_{\rm y}}\right)\exp\left\{\jmath2\pi t\nu_{k,l}\right\}\notag\\
\exp\left\{-\jmath2\pi f_c\tau_{n_{\rm x},n_{\rm y}}\left(\bm{\vartheta}_k\right)\right\}.
\end{align}
By applying the Fourier transform to \eqref{eq:httau}, the complex baseband spatial response of the downlink channel between the $(n_{\rm x},n_{\rm y})$th antenna and the $k$th UT at instant $t$ with frequency $f$ can be expressed as \cite{you2020massive}
\begin{align}\label{eq:cbsr}
h_{k,n_{\rm x},n_{\rm y}}(t,f)=\sum_{l=1}^{L_k}&\alpha_{k,l}\exp\left\{\jmath2\pi\left[ t\nu_{k,l}-f\tau_{k,l}\right]\right\}\notag\\
&\exp\left\{-\jmath2\pi (f_c+f)\tau_{n_{\rm x},n_{\rm y}}\left(\bm{\vartheta}_k\right)\right\},
\end{align}
which holds with invariant parameters $L_k$, $\alpha_{k,l}$, $\tau_{k,l}$, $\tau_{n_{\rm x},n_{\rm y}}\left(\bm{\vartheta}_k\right)$, $\nu_{k,l}$ during the time intervals of interest \cite{you2020massive}. Note that the update of these parameters should be performed when the positions of the UTs and LEO satellite change significantly.

We denote $\tau_k^{\rm min}$ as the minimum propagation delay for the $k$th UT, i.e., $\tau_k^{\rm min}=\min_{l}\left\{\tau_{k,l}\right\}$ and the downlink channel in \eqref{eq:cbsr} can be reorganized as
\begin{align}\label{eq:madc}
\bH_k\left(t,f\right)=\exp\{\jmath 2\pi[t\nu_k^{\rm sat}-f\tau_k^{\rm min}]\}g_k(t,f)\bm{\Psi}_k(f),
\end{align}
where the $\left(n_{\rm x},n_{\rm y}\right)$th element in $\bm{\Psi}_k(f)\in \mathbb{C}^{N^{\rm x}_{\rm t}\times N^{\rm y}_{\rm t}}$ is expressed as $\exp\left\{-\jmath2\pi (f_c+f)\tau_{n_{\rm x},n_{\rm y}}\left(\bm{\vartheta}_k\right)\right\}$, and the channel gain $g_k(t,f)$ is defined as 
\begin{align}
g_k(t,f)\triangleq \sum_{l=1}^{L_k}\alpha_{k,l}\exp\{\jmath 2\pi[t(\nu_{k,l}-\nu_k^{\rm sat})-f(\tau_{k,l}-\tau_k^{\rm min})]\}.
\end{align}
In this work, the channel gain is modeled as a Rician distributed random variable with Rician factor $\kappa_k$ and power $\mathbb{E}\{|g_k(t,f)|^2\}=\gamma_k$ \cite{you2020massive}.
Then, for notation brevity, we rewritten $\bH_k\left(t,f\right)$ into a vector form, given by
\begin{align}\label{eq:orgcmr}
\bh_k\left(t,f\right)=\exp\{\jmath 2\pi[t\nu_k^{\rm sat}-f\tau_k^{\rm min}]\}g_k(t,f)\bv_k(f),
\end{align}
where $\bv_k\left(f\right)$ is the frequency-dependent UPA response vector considering the beam squint effects, given by
    \begin{align}\label{eq:arrp}
    \mathbf{v}_k(f)&\triangleq\mathbf{v}(f,\bm{\vartheta}_k)=\mathbf{v}_k^{\mathrm{x}}(f)\otimes \mathbf{v}_k^{\mathrm{y}}(f)\notag\\
    &=\mathbf{v}_{\mathrm{x}}\left(f,\vartheta_k^{\mathrm{x}}\right)\otimes\mathbf{v}_{\mathrm{y}}\left(f,\vartheta_k^{\mathrm{y}}\right)\in \mathbb{C}^{N_{\rm t}\times 1}.
    \end{align}
The array response vector $\mathbf{v}_k^d(f)\in\mathbb{C}^{N_\mathrm{t}^d\times 1}$ for $d\in \mathcal{D}\triangleq \{\rm x,y\}$ is defined as \cite{you2020massive}
    \begin{align}\label{eq:crvc}
    \mathbf{v}_k^d(f)&\triangleq \mathbf{v}_d\left(f,\vartheta_k^d\right)\in \bbC^{N_{\rm t}^d\times 1}\notag\\
    &=\frac{1}{\sqrt{N_\mathrm{t}^d}}\left[1\ \exp\left\{-\jmath\phi(f,\vartheta_k^d)\right\}\ \cdots\ \right.\notag\\ &\qquad \qquad \qquad \left.\exp\left\{-\jmath\phi(f,\vartheta_k^d)(N_\mathrm{t}^d-1)\right\}\right]^T,
    \end{align}
where $\displaystyle \phi\left(f,\vartheta_k^d\right)\triangleq 2\pi(f_c+f)\frac{r}{c}\vartheta_k^d$.
Subsequently, with proper time and frequency synchronization \cite{you2020massive},
the complex-valued baseband downlink space-frequency channel response for the $k$th UT can be equivalently expressed as $\mathbf{h}_k(t,f)=\mathbf{v}_k(f)g_k(t,f)$ \cite{you2020massive,dovelos2021channel}.

In the following, we focus on each coherence time interval and omit the index $t$ for simplicity.
Besides, at the $m$th subcarrier with frequency $f_m$, we denote $\bh_k[m]\triangleq \bh_k(f_m)$, $\bv_k[m]\triangleq \bv_k(f_m)$, and $g_k[m]\triangleq g_k(f_m)$. Then, the channel response vector at the $m$th subcarrier can be expressed as $\mathbf{h}_k[m]=\mathbf{v}_k[m]g_k[m]$.
Note that massive MIMO is adopted in this work with a large UPA array, leading to high time delay according to \eqref{eq:tmdl} and large variation for the UPA array response across the OFDM subcarriers according to \eqref{eq:arrp} and \eqref{eq:crvc}, especially for the considered wide-band systems.
According to Eq. \eqref{eq:ofdmf}, the OFDM frequency $f_m$ is inversely proportional to the signal duration $T_{\rm s}$. Then, the corresponding UPA response $\bv_k[m]$ not only depends on the AoD information, but also the ratio of the time delay $\tau_{n_{\rm x},n_{\rm y}}\left(\bm{\vartheta}_k\right)$ and $T_{\rm s}$ \cite{wang2018spatial}.
When the signal duration $T_{\rm s}$ is comparable to
or much smaller than the time delay $\tau_{n_{\rm x},n_{\rm y}}\left(\bm{\vartheta}_k\right)$, the influence of this ratio on the array response can not be ignored, and the beam squint effects appear \cite{dovelos2021channel,wang2018spatial}.
\subsection{Downlink Transmission Signal Model in the Communication Module}
The transmitter at the satellite side contains a hybrid precoder with $M_{\rm t}$ ($K\leq M_{\rm t}\leq N_{\rm t}$) RF chains and each UT is equipped with a fully digital architecture.
At the $m$th subcarrier, the final transmit signal vector is denoted by $\bx[m]=\bB[m]\bs[m] \in \mathbb{C}^{N_{\rm t}\times 1}$ where $\bs[m]=\left[s_1[m],s_2[m],\ldots,s_K[m]\right]^T \in \mathbb{C}^{K\times 1}$ is the transmit data vector with zero mean and autocorrelation matrix $\mathbb{E}\{\bs[m]\bs^H[m]\}=\bI_K, \ \forall m$.
The hybrid precoder $\bB[m]$ consists of a baseband precoder $\mathbf{W}_{\rm BB}[m]=\left[\mathbf{w}_{{\rm BB},1}[m],\mathbf{w}_{{\rm BB},2}[m],\ldots,\mathbf{w}_{{\rm BB},K}[m]\right]\in \mathbb{C}^{M_\mathrm{t}\times K}$ and a unit-modulus frequency-dependent\footnote{The frequency-dependent RF analog precoder can be implemented via, e.g., the combination of the frequency-independent phase-shifting network and several true-time-delay elements \cite{dai2021delay}, and the time-delay phase shifting network \cite{lin2017subarray}.} RF analog precoder $\bW_{\rm RF}[m]\in \mathbb{C}^{N_\mathrm{t}\times M_\mathrm{t}}$. Then, it can be expressed as $\bB[m]=\bW_{\rm RF}[m]\bW_{\rm BB}[m]=\left[\bb_1[m],\bb_2[m],\ldots,\bb_K[m]\right]\in \mathbb{C}^{N_{\rm t}\times K}$ and the precoding vector for the $k$th UT is given by, $\bb_k[m]=\bW_{\rm RF}[m]\bw_{{\rm BB},k}[m]\in \mathbb{C}^{N_{\rm t}\times 1}$.

Based on the above model, the received signal of UT $k$ at the $m$th subcarrier is given by
    \begin{align}\label{eq:rcsg}
y_k[m]&=\mathbf{h}_k^H[m]\bx[m]+n_k[m]\notag\\
&=\mathbf{h}_k^H[m]\mathbf{b}_k[m]s_k[m]\notag\\
&\qquad \qquad +\mathbf{h}_k^H[m]\sum_{i\neq k}\mathbf{b}_i[m]s_i[m]+n_k[m],
    \end{align}
where the additive Gaussian white noise $n_k[m]$ is distributed as $n_k\sim \mathcal{CN}(0,N_0^{\rm c})$. Then, the corresponding signal-to-interference-plus-noise ratio (SINR) at the downlink between the satellite and the $k$th UT at the $m$th subcarrier is given by
    \begin{align}\label{eq:dlsinr}
    \sinr_{k}[m]\triangleq \frac{|\mathbf{b}_k^H[m]\mathbf{h}_k[m]|^2}{\sum_{\ell\neq k}|\mathbf{b}_{\ell}^H[m]\mathbf{h}_k[m]|^2+N_0^{\rm c}}.
    \end{align}
Note that the downlink channel vector $\bh_k[m], \forall k,m$ is difficult to estimate at the transmitter of the LEO satellite in the considered communication module, due to the characteristics of the long propagation delay as well as the mobility of both the UTs and satellites, as mentioned in the Introduction Section.
Hence, in this work, the precoder for the downlink of the LEO satellite ISAC system is designed based on the sCSI knowledge, which is based on the statistical information of the channel gain $g_k[m],\forall k$ and the UPA response vector $\bv_k[m], \forall k, m$.

According to the expression of the SINR in \eqref{eq:dlsinr}, the ergodic rate for the $k$th UT at the $m$th subcarrier is given by 
    \begin{align}\label{eq:rate}
    R_k=\sum_{m=1}^M\Delta_BR_k[m]=\sum_{m=1}^M\Delta_B\mathbb{E}\{\log(1 + \sinr_k[m])\},
    \end{align}
and the EE for the communication module is defined as 
    \begin{align}\label{eq:eedf}
    {\rm EE}=\frac{\sum_{k=1}^KR_k}{P_{\mathrm{total}}}.
    \end{align}
Note that the total transmission power consumption $P_{\mathrm{total}}$ is given by 
    \begin{equation}\label{eq:totp}
    \begin{aligned}
    P_{\mathrm{total}}=\sum_{k=1}^K\sum_{m=1}^M\xi ||\bb_k[m]||_2^2+P_\mathrm{t},
    \end{aligned}
    \end{equation}
where $1/\xi$ is the efficiency of the power amplifier and $P_{\rm t}$ denotes the static power consumed by the hybrid precoder, defined as \cite{mendezrial2016hybrid, chen2018low}
    \begin{align}\label{eq:fppt}
    P_{\rm t}=M_\mathrm{t}P_{\mathrm{RFC}}+P_{\mathrm{LO}}+P_{\mathrm{BB}}+P_{\rm AL}.
    \end{align}
In particular, $P_{\mathrm{RFC}}$, $P_{\mathrm{LO}}$, $P_{\mathrm{BB}}$ denote the power consumption of one RF chain, the local oscillator, and the baseband digital precoder, respectively.
The power consumed by the RF analog precoder is represented by $P_{\rm AL}$, which is mainly related to the number of antennas and transmission bandwidth \cite{lin2017subarray,li2018beam}.
\subsection{Sensing Beampattern Considering Beam Squint Effects}
The transmit beampattern for the sensing module at the $m$th subcarrier is given by \cite{li2008mimo,fan2019mimo,liu2019hybrid,ahmed2019ofdm,liu2020joint,kaushik2021hardware,cheng2021hybrid}
\begin{align}\label{eq:radb}
Q_m(\bm{\vartheta})=\bv_m^H(\bm{\vartheta})\bX[m]\bv_m(\bm{\vartheta}), \forall \bm{\vartheta},
\end{align}
where $\bv_m^H(\bm{\vartheta})\triangleq \bv(f_m,\bm{\vartheta})$, $\bm{\vartheta}=\left(\vartheta^{\rm x},\vartheta^{\rm y}\right)$ denotes the space angle pair, and $\bX[m]$ is the covariance matrix, defined as \cite{xu2021wideband} 
    \begin{align}\label{eq:covx}
    \bX[m]&=\mathbb{E}\{\bx[m]\bx^H[m]\}\notag\\
    &=\bW_{\rm RF}[m]\bW_{\rm BB}[m]\bW_{\rm BB}^H[m]\bW_{\rm RF}^H[m].
    \end{align}

Assuming that there exist $P_{\rm r}\leq K$ targets, then the optimal sensing precoder at the $m$th subcarrier is given by \cite{yoon2006tops,hassanien2010phased,liu2019hybrid,kaushik2021hardware}
\begin{align}\label{eq:optbrad}
\bB_{\rm ss}[m]=\blkdiag{\bu_1[m],\bu_2[m],\ldots,\bu_{P_{\rm r}}[m]}
\in \bbC^{N_{\rm t}\times P_{\rm r}},
\end{align}
where the diagonal vector $\bu_p[m]\in \mathbb{C}^{N_{\rm t}/P_{\rm r}\times 1}$ denotes the corresponding $N_{\rm t}/P_{\rm r}$ elements of the UPA array response $\bv_p[m]$ for all $p\in\{1,2,\cdots,P_{\rm r}\}$. Note that, due to the beam squint effects, the sensing beampattern is not only determined by the AoD information, but also depends on the OFDM subcarrier frequencies \cite{yoon2006tops}.
\subsection{Probability of Muti-Target Detection}
The satellite first transmits omnidirectional beams to probe the potential targets in any direction \cite{liu2020joint}. Then, the satellite estimates the AoD information of the targets and formulates directional downlink waveforms with optimized ISAC precoders towards the targets of interest, to achieve better observation of the targets \cite{li2008mimo}.
In particular, at the $m$th subcarrier, the reflected signal is given by \cite{liu2020joint}
\begin{align}\label{eq:radecho}
\by_{\rm r}[m]=\sum_{p=1}^{P_{\rm r}}\beta_{p}\bv_m(\bm{\vartheta}_{p})\bv_m^T(\bm{\vartheta}_{p})\bx_{\rm r}[m]+\bz[m] \in \mathbb{C}^{N_{\rm t}\times 1},
\end{align}
where $P_{\rm r}$ is the number of the targets, $\beta_{p}$ denotes the reflection coefficient of the $p$th target, $\bv_m(\bm{\vartheta}_p)$ is the array response vector with respect to the space angle pair $\bm{\vartheta}_p=\left(\vartheta_p^{\rm x},\vartheta_p^{\rm y}\right)$ for target $p$, and $\bz[m]$ is the additive white Gaussian
noise with variance $N_0^{\rm r}$.
Besides, $\bx_{\rm r}[m]$ is the probing signal, satisfying $\mathbb{E}\left\{\bx_{\rm r}[m]\bx_{\rm r}[m]^H\right\}=\frac{P}{MN_{\rm t}}\bI_{N_{\rm t}}, \forall m$, where $P$ denotes the transmit power budget \cite{liu2020joint}.
For mathematical convenience, we denote $\bA\left(\Theta_m\right)=\left[\bv_m(\bm{\vartheta}_{1}),\cdots,\bv_m(\bm{\vartheta}_{P_{\rm r}})\right]$ and $\diag{\bm{\beta}}=\diag{\beta_1,\cdots,\beta_{P_{\rm r}}}$.
Subsequently, Eq. \eqref{eq:radecho} can be rewritten as
\begin{align}
\by_{\rm r}[m]=\bA\left(\Theta_m\right)\diag{\bm{\beta}}\bA^T\left(\Theta_m\right)\bx_{\rm r}[m]+\bz[m].
\end{align}

Then, according to \cite{khawar2015target}, the probability of detection is calculated as
\begin{align}\label{eq:pod}
P_{\rm D}=1-\mathcal{F}_{\mathcal{X}_{2P_{\rm r}}^2(\varsigma)}\left(\mathcal{F}^{-1}_{\mathcal{X}_{2P_{\rm r}}^2}\left(1-P_{\rm FA}\right)\right),
\end{align}
where $P_{\rm FA}$ denotes the false alarm probability.
In addition, $\mathcal{F}^{-1}_{\mathcal{X}_{2P_{\rm r}}^2}$ is the inverse central chi-squared distribution
with degrees of freedom being $2P_{\rm r}$ and $\mathcal{F}_{\mathcal{X}_{2P_{\rm r}}^2(\varsigma)}$
is the noncentral chi-squared distribution with $2P_{\rm r}$ degrees of freedom and noncentrality parameter
given by \cite{khawar2015target}
\begin{align}
\varsigma=\frac{\sum_{m=1}^M\mathbb{E}\left\{\abs{\abs{\bA\left(\Theta_m\right)\diag{\bm{\beta}}\bA^T\left(\Theta_m\right)\bx_{\rm r}[m]}}_2^2\right\}}{MN_0^{\rm r}}.
\end{align}
\subsection{Problem Formulation}
The objective of our work is to design a beam squint-aware hybrid precoder for the considered wide-band downlink massive MIMO LEO satellite ISAC systems.
Note that the corresponding optimization problem has multiple objectives and is generally not easy to handle.
To that end, we formulate the following optimization problem, to maximize the EE of communications while guaranteeing the sensing performance, as follows
    \begin{subequations}\label{eq:eemn}
    \begin{align}
    \mathcal{P}_1:\mathop{\mathrm{maximize}}\limits_{\substack{\{\bW_{\rm RF}[m],\\\bW_{\rm BB}[m],\\ \bU[m]\}_{m=1}^M}}&\ \ \frac{\sum_{k=1}^K R_k}{P^{\mathrm{total}}} \label{eq:eemna}\\
    \mathrm{s.t.}\ \ \ &\ \ \sum_{k=1}^K\sum_{m=1}^M||\bW_{\rm RF}[m]\bw_{{\rm BB},k}[m]||_2^2\leq P,\label{eq:eemnb}\\
    \ \ \ \ &\ \ \bW_{\rm RF}[m]\in \mathcal{S},\ \forall m, \label{eq:eemnc}\\
    \ \ \ \ &\ \ \abs{\abs{\bW_{\rm RF}[m]\bW_{\rm BB}[m]-\bB_{\rm ss}[m]\bU[m]}}_F^2\leq \varepsilon,\ \notag\\&\qquad \qquad \qquad \qquad \qquad \qquad \qquad \forall m, \label{eq:eemnd}\\
    \ \ \ \ &\ \ \bU[m]\bU^H[m]=\bI_{P_{\rm r}},\ \forall m, \label{eq:eemne}
    \end{align}
    \end{subequations}
where an auxiliary unitary matrix $\bU[m]\in \bbC^{P_{\rm r}\times K}$ at the $m$th subcarrier is introduced to match the dimensionality with the product of the hybrid precoders, since its multiplication with $\bB_{\rm ss}[m]$ has no effects on the sensing beampattern \cite{kaushik2021hardware}.
Besides, $\varepsilon$ denotes the tolerance item of the Euclidean distance between the hybrid digital/analog precoders and the sensing precoder (with rotation).
Note that the RF analog precoder can be designed by following either a fully or a partially connected structure \cite{zhu2019millimeter,arora2019hybrid}.
In particular, every antenna element is connected to all the RF chains in the fully connected structure, while the antennas are split into $M_{\rm t}$ groups and each group involves $N_{\rm g}=N_{\rm t}/M_{\rm t}$ elements connected to the same RF chain in the partially connected case.
Then, the non-zero elements of the RF analog precoder $\bW_{\rm RF}[m]$ at the $m$th subcarrier are selected from the set $\cS \triangleq \left\{\cS_{\rm FC},\cS_{\rm PC}\right\}$, given by
\begin{align}\label{eq:vconstr}
\mathcal{S}_{\rm FC}&\triangleq\left\{\bW_{\rm RF}\Big |\abs{\left[\bW_{\rm RF}\right]_{i,j}}=1,\ \forall i,j \right\},\\
\mathcal{S}_{\rm PC}&\triangleq\left\{\bW_{\rm RF}\Big |\abs{\left[\bW_{\rm RF}\right]_{i,j}}=1,\ \forall i, \ \forall j=\left\lceil \frac{i}{N_{\rm g}}\right\rceil\right\},
\end{align}
for the fully and partially connected structures, respectively \cite{arora2019hybrid}.

The problem in \eqref{eq:eemn} involves a fractional objective and several nonconvex constraints.
In addition, the ergodic rate in the numerator and the total power consumption in the denominator of the objective function are both nonconvex with regard to the two tightly coupled variables, i.e., the baseband digital and the RF analog precoders.
Thus, the problem is, in general, difficult to solve.
To tackle the problem, we first consider the product of the baseband digital and the RF analog precoders at each subcarrier as an equivalent digital precoder, and omit the unrelated constraints temporarily \cite{zi2016energy,kaushik2021hardware}.
With the obtained fully digital precoder, we introduce a weighting factor to balance the performance of communications and sensing.
Then, the corresponding problem can be tackled through an alternating optimization framework \cite{zi2016energy,kaushik2021hardware}.
\section{Beam Squint-Aware Equivalent Fully Digital Precoding}\label{sec:bs_fdp}
In this section, we focus on the equivalent fully digital problem, which is first converted into several subproblems via Dinkelbach's algorithm. Then, with the Lagrangian dual transform and the quadratic transform, each subproblem can be transformed into a convex one, which can be handled through a Lagrange multiplier method.
\subsection{Upper Bound of the Ergodic Rate}
One of the difficulties to tackle problem $\cP_1$ lies in the expectation operator of the ergodic rate in the numerator of the objective function. The Monte-Carlo method can be employed to estimate the results of the expectation operation but with high computational complexity \cite{li2020downlink}. Therefore, in this work, we substitute the ergodic rate with its tight upper bound. Specifically, based on \cite[Lemma 2]{sun2015beam}, the rate expression, $\log(1 + \sinr_k[m])$, can be shown a concave function with respect to the channel gain $g_k$. Then, Jensen's inequality can be adopted to derive the upper bound of each ergodic rate $R_k[m]$, given by
    \begin{align}\label{eq:exrt}
    R_k[m]&\leq \bar{R}_k[m]\notag\\
    &=\log\left(1+\frac{\gamma_k|\mathbf{v}_k^H[m]\mathbf{b}_k[m]|^2}{\sum_{{\ell}\neq k}\gamma_k|\mathbf{v}_k^H[m]\mathbf{b}_{\ell}[m]|^2+N_0^{\rm c}}\right).
    \end{align}
    \emph{Proof:} Please refer to \appref{app:a}.
\subsection{Fractional Programming}
The fully digital equivalent problem is a fractional one with the numerator and denominator of the objective being continuous and positive-valued.
Thus, it can be tackled first via Dinkelbach's algorithm \cite{zapponealessio2015energy,rodenas1999extensions}.
In particular, with indexed variable $\eta^{\left(i\right)},\ i=1,2,\ldots$, a series of auxiliary subproblems, whose solutions are guaranteed to converge to the globally optimal point of the original fully digital equivalent problem \cite{zapponealessio2015energy}, are introduced.
Then, we denote $\mathcal{B}^{\left(i\right)}=\{\bB^{\left(i\right)}[m]\}_{m=1}^M$ and the $i$th subproblem is given by \cite{zapponealessio2015energy,rodenas1999extensions}
    \begin{subequations}\label{eq:spdb}
    \begin{align}
    \mathcal{P}_2^{\left(i\right)}: \mathop{\mathrm{maximize}}\limits_{\cB^{\left(i\right)}, \eta^{\left(i\right)}}&\ \ F(\cB^{\left(i\right)},\eta^{\left(i\right)})=\sum_{k=1}^K\bar{R}_k\left(\cB^{\left(i\right)}\right)\notag\\
    &\qquad \qquad \qquad \quad -\eta^{\left(i\right)} P^{\mathrm{total}}\left(\cB^{\left(i\right)}\right)\label{eq:dbso}\\
    \mathrm{s.t.}&\ \ \sum_{k=1}^K\sum_{m=1}^M||\bb_k^{\left(i\right)}[m]||_2^2\leq P.\label{eq:dbsc}
    \end{align}
    \end{subequations}
In the following, an alternating optimization framework is adopted to handle the subproblems by iteratively optimizing the parameters $\cB^{\left(i\right)}$ and $\eta^{\left(i\right)}$.
In particular, with given $\cB^{\left(i\right)}$, the auxiliary variable $\eta^{\left(i\right)}$ is updated as \cite{zapponealessio2015energy,rodenas1999extensions}
    \begin{align}
\eta^{\left(i+1\right)}=\frac{\sum_{k=1}^K\bar{R}_k\left(\cB^{\left(i\right)}\right)}{P^{\mathrm{total}}\left(\cB^{\left(i\right)}\right)}.
    \end{align}
Subsequently, we focus on the optimization of $\cB^{\left(i\right)}$ with fixed $\eta^{\left(i\right)}$ and the index $i$ is omitted in the following for brevity. Let $\bb_k[m]=\bb_{k,m}$ and $\bv_k[m]=\bv_{k,m}$, we can rewrite problem \eqref{eq:spdb} as
    \begin{subequations}
    \begin{align}
    \mathcal{P}_2: \mathop{\mathrm{maximize}}\limits_{\cB}&\ \ F(\mathcal{B})\label{eq:dbof}\\
    \mathrm{s.t.}&\ \ \sum_{k=1}^K\sum_{m=1}^M||\bb_{k,m}||_2^2\leq P,\label{eq:dbpc}
    \end{align}
    \end{subequations}
where the objective function is shown in Eq. \eqref{eq:obj1} on the top of the next page.
\newcounter{TempEqCnt1} 
\setcounter{TempEqCnt1}{\value{equation}} 
\setcounter{equation}{31} 
\begin{figure*}[ht] 
    \begin{align}\label{eq:obj1}
    F(\mathcal{B})=\sum_{k=1}^K\sum_{m=1}^M\log\left(1+\frac{\gamma_k|\mathbf{v}_{k,m}^H\mathbf{b}_{k,m}|^2}{\sum_{{\ell}\neq k}\gamma_k|\mathbf{v}_{k,m}^H\mathbf{b}_{\ell,m}|^2+N_0^{\rm c}}\right)-\eta \left(\sum_{k=1}^K\sum_{m=1}^M\xi||\bb_{k,m}||_2^2+P_\mathrm{t}\right)
    \end{align}
    \hrule
\end{figure*}

Note that after the utilization of Dinkelbach's algorithm, problem $\cP_2$ still involves a sum-of-logarithms term with a fractional expression in each logarithmic operator, which is not easy to handle.
Hence, according to \cite{shen2018fractional}, the Lagrangian dual transform is adopted to substitute the fractional expressions with a corresponding parameter from an auxiliary variable collection $\lambda=\left\{\lambda_{k,m}\right\}_{k=1,m=1}^{K,M}$, and settle the ratio in \eqref{eq:dbof} outside the logarithmic operator.
Then, problem $\cP_2$ is converted into \cite{shen2018fractional}
    \begin{subequations}
    \begin{align}
    \cP_3: \mathop{\mathrm{maximize}}\limits_{\cB, \lambda}&\ \ F(\cB, \lambda)\label{eq:lelo}\\
    \mathrm{s.t.}&\ \ \sum_{k=1}^K\sum_{m=1}^M||\bb_{k,m}||_2^2\leq P,\label{eq:lelc}
    \end{align}
    \end{subequations}
where the objective function is presented in Eq. \eqref{eq:obj2} on the top of the next page.
\newcounter{TempEqCnt2} 
\setcounter{TempEqCnt2}{\value{equation}} 
\setcounter{equation}{33} 
\begin{figure*}[ht] 
    \begin{align}\label{eq:obj2}
    F(\cB, \lambda)=\sum_{k=1}^K\sum_{m=1}^M\log\left(1+\lambda_{k,m}\right)+\left(1+\lambda_{k,m}\right)\frac{\gamma_k|\mathbf{v}_{k,m}^H\mathbf{b}_{k,m}|^2}{\sum_{{\ell}=1}^K\gamma_k|\mathbf{v}_{k,m}^H\mathbf{b}_{\ell,m}|^2+N_0^{\rm c}}-\lambda_{k,m} -\eta \left(\sum_{k=1}^K\sum_{m=1}^M\xi||\bb_{k,m}||_2^2+P_\mathrm{t}\right)
    \end{align}
    \hrule
\end{figure*}
It is worth noting that for fixed $\cB$, problem $\cP_3$ is concave with respect to $\lambda$. Then, let $\partial F/\partial \lambda_{k,m}$ equal to zero, we obtain the optimal update of each $\lambda_{k,m}$, which is given by \cite{shen2018fractional}
\begin{align}\label{eq:optlambda}
\lambda_{k,m}^{\rm opt}=\frac{\gamma_k|\mathbf{v}_{k,m}^H\mathbf{b}_{k,m}|^2}{\sum_{{\ell}\neq k}\gamma_k|\mathbf{v}_{k,m}^H\mathbf{b}_{\ell,m}|^2+N_0^{\rm c}}.
\end{align}
It is worth noting that by substituting \eqref{eq:optlambda} into \eqref{eq:lelo}, we can recover the objective function in \eqref{eq:dbof}.
Then, $\cB$ is the solution of problem $\cP_2$ if and only if it can solve problem $\cP_3$, and the corresponding optimal values of the objectives in $\cP_2$ and $\cP_3$ are the same.
The equivalence between problems $\cP_2$ and $\cP_3$ is therefore established.

Next, regarding the sum-of-ratios term in problem $\cP_3$, we apply the quadratic transform \cite{shen2018fractional} with another auxiliary variable $\rho=\left\{\rho_{k,m}\right\}_{k=1,m=1}^{K,M}$ introduced and obtain the following equivalent problem \cite{shen2018fractional}
    \begin{subequations}
    \begin{align}
    \cP_4: \mathop{\mathrm{maximize}}\limits_{\cB, \lambda,\rho}&\ \ F(\cB,\lambda,\rho)\label{eq:qtof}\\
    \mathrm{s.t.}&\ \ \sum_{k=1}^K\sum_{m=1}^M||\bb_{k,m}||_2^2\leq P,\label{eq:qtpc}
    \end{align}
    \end{subequations}
where the objective function is given in Eq. \eqref{eq:obj3} on top of the next page.
\newcounter{TempEqCnt3} 
\setcounter{TempEqCnt3}{\value{equation}} 
\setcounter{equation}{36} 
\begin{figure*}[ht] 
    \begin{align}\label{eq:obj3}
    F(\cB,\lambda,\rho)=\sum_{k=1}^K\sum_{m=1}^M\log\left(1+\lambda_{k,m}\right)-\lambda_{k,m}+2\sqrt{\left(1+\lambda_{k,m}\right)\gamma_k}\Re\left\{\bb_{k,m}^H\bv_{k,m}\rho_{k,m}\right\}\notag\\
-\abs{\rho_{k,m}}^2\left(\sum_{{\ell}=1}^K\gamma_k|\mathbf{v}_{k,m}^H\mathbf{b}_{\ell,m}|^2+N_0^{\rm c}\right)-\eta \left(\sum_{k=1}^K\sum_{m=1}^M\xi||\bb_{k,m}||_2^2+P_\mathrm{t}\right)
    \end{align}
    \hrule
\end{figure*}
Similarly, the optimal value of each auxiliary $\rho_{k,m}$ can be determined by setting $\partial F/\partial \rho_{k,m}$ equal to zero with all the other variables fixed, which is given by
\begin{align}
\rho_{k,m}^{\rm opt}=\frac{\sqrt{\left(1+\lambda_{k,m}\right)\gamma_k}\mathbf{v}_{k,m}^H\mathbf{b}_{k,m}}{\sum_{{\ell}=1}^K\gamma_k|\mathbf{v}_{k,m}^H\mathbf{b}_{\ell,m}|^2+N_0^{\rm c}}.
\end{align}

It is not difficult to confirm that for fixed $(\eta,\lambda,\rho)$, problem $\cP_4$ is convex with respect to $\bb_{k,m}$. To efficiently solve problem $\cP_4$, we attach a Lagrange multiplier $t$ to the power constraint \eqref{eq:qtpc} and the corresponding Lagrange function is given by
    \begin{equation}\label{eq:lfun}
    \begin{aligned}
    L(\cB,\lambda,\rho,t)
    =F(\cB,\lambda,\rho)+t\left(\sum_{k=1}^K\sum_{m=1}^M||\bb_{k,m}||_2^2-P\right).
    \end{aligned}
    \end{equation}

Subsequently, the precoding vector towards the $k$th UT at the $m$th subcarrier can be determined by setting the derivative of the Lagrange function $L(\cB,\lambda,\rho,t)$ equal to zero, i.e., $\partial L/\partial \bb_{k,m}=\mathbf{0}$, and the result is derived as follows \cite{shi2011iteratively}
\begin{align}\label{eq:bopt}
\bb_{k,m}^{\rm opt}=\left(\sum_{\ell=1}^K\abs{\rho_{\ell,m}}^2\gamma_{\ell}\bv_{\ell,m}\bv_{\ell,m}^H+\left(\eta\xi+t\right)\bI\right)^{-1}\notag\\
\sqrt{\left(1+\lambda_{k,m}\right)\gamma_k}\rho_{k,m}\bv_{k,m},
\end{align}
where the Lagrange multiplier $t$ is selected to satisfy the following complementarity slackness condition
\begin{equation}
\begin{aligned}
t=\arg\ \min_{t\geq 0}\ \ \sum_{k=1}^K\sum_{m=1}^M||\bb_{k,m}^{\rm opt}||_2^2\leq P.
\end{aligned}
\end{equation}
Moreover, the optimal value of the Lagrange multiplier $t$ can be calculated with the following steps. First, let $\bb_{k,m}\left(t\right)$ denotes the right hand side of \eqref{eq:bopt}, $\mathbf{\Psi}_m=\sum_{\ell=1}^K\abs{\rho_{\ell,m}}^2\gamma_{\ell}\bv_{\ell,m}\bv_{\ell,m}^H$ and $\bA_m=\mathbf{\Psi}_m+\eta\xi\bI$.
Note that $\mathbf{\Psi}_m$ is a positive semi-definite Hermitian matrix, and thus, $\bA_m$ is invertible for $\eta\neq 0$. Then, if $\bA_m$ is invertible and $\sum_{k=1}^K\sum_{m=1}^M||\bb_{k,m}\left(0\right)||_2^2\leq P$, the equation $\bb_{k,m}^{\rm opt}=\bb_{k,m}\left(0\right)$ holds, otherwise the following condition must be tackled
\begin{align}\label{eq:pcstr}
\sum_{k=1}^K\sum_{m=1}^M{\rm Tr}\left(\bb_{k,m}\left(t\right)\bb_{k,m}^H\left(t\right)\right)=P,
\end{align}
which can be cast in an easier-to-handle form. Note that $\bA_m$ is a Hermitian matrix, which admits the eigenvalue decomposition given by $\bA_m=\bD_m\mathbf{\Lambda}_m\bD_m^H$. Hence, \eqref{eq:pcstr} can be equivalently converted into
\begin{align}\label{eq:edec}
\sum_{m=1}^M{\rm Tr}\left(\left(\mathbf{\Lambda}_m+t\bI\right)^{-2}\mathbf{\Phi}_m\right)=P,
\end{align}
where $\mathbf{\Phi}_m=\bD_m^H\left(\sum_{k=1}^K\left(1+\lambda_{k,m}\right)\gamma_k\abs{\rho_{k,m}}^2\bv_{k,m}\bv_{k,m}^H\right)\bD_m$. Following the properties the trace operator, the above Eq. \eqref{eq:edec} can be further simplified as
\begin{align}\label{eq:dett}
\sum_{m=1}^M\sum_{n=1}^{N_{\rm t}}\frac{\left[\mathbf{\Phi}_m\right]_{n,n}}{\left(\left[\mathbf{\Lambda}_m\right]_{n,n}+t\right)^2}=P,
\end{align}
where $t$ can be determined through the bisection method \cite{boyd2004convex}.
We summarize the procedure for designing the beam squint-aware equivalent fully digital precoder in \textbf{Algorithm \ref{alg:algdasam}}.

\begin{algorithm}[t!]
\caption{Beam Squint-Aware Equivalent Fully Digital Precoding}
\label{alg:algdasam}
\begin{algorithmic}[1]
\REQUIRE Threshold $\epsilon_1>0$, $i=0$, $\eta^{\left(0\right)}=0$
\ENSURE Equivalent fully digital precoder $\bB\left[m\right],\ m=1,2,\cdots, M$
\STATE Initialize $\cB^{\left(0\right)}=\left\{\bb_{k,m}^{\left(0\right)}\right\}_{k=1,m=1}^{K,M}$ such that $\sum_{k=1}^K\sum_{m=1}^M||\bb_{k,m}^{\left(0\right)}||_2^2= P$
\WHILE{$F(\cB^{\left(i\right)},\eta^{\left(i\right)})>\epsilon_1$}
\REPEAT
\STATE Update corresponding auxiliary variables and the precoding vector as
\begin{align}
\lambda_{k,m}^{\rm opt}&=\frac{\gamma_k|\mathbf{v}_{k,m}^H\mathbf{b}_{k,m}|^2}{\sum_{{\ell}\neq k}\gamma_k|\mathbf{v}_{k,m}^H\mathbf{b}_{\ell,m}|^2+N_0^{\rm c}}, \ \forall k,m,\\
\rho_{k,m}^{\rm opt}&=\frac{\sqrt{\left(1+\lambda_{k,m}\right)\gamma_k}\mathbf{v}_{k,m}^H\mathbf{b}_{k,m}}{\sum_{{\ell}=1}^K\gamma_k|\mathbf{v}_{k,m}^H\mathbf{b}_{\ell,m}|^2+N_0^{\rm c}}, \ \forall k,m, \\
\bb_{k,m}^{\rm opt}&=\left(\sum_{\ell=1}^K\abs{\rho_{\ell,m}}^2\gamma_{\ell}\bv_{\ell,m}\bv_{\ell,m}^H+\left(\eta\xi+t\right)\bI\right)^{-1}\notag\\
&\qquad \sqrt{\left(1+\lambda_{k,m}\right)\gamma_k}\rho_{k,m}\bv_{k,m}, \ \forall k,m,
\end{align}
where $t$ is determined according to Eq. \eqref{eq:dett}
\UNTIL The convergence of objective \eqref{eq:lelo}
\STATE $F(\cB^{\left(i\right)},\eta^{\left(i\right)})=\sum_{k=1}^K\bar{R}_k\left(\cB^{\left(i\right)}\right)-\eta^{\left(i\right)} P^{\mathrm{total}}\left(\cB^{\left(i\right)}\right)$
\STATE $\eta^{\left(i+1\right)}=\sum_{k=1}^K\bar{R}_k\left(\cB^{\left(i\right)}\right)/P^{\mathrm{total}}\left(\cB^{\left(i\right)}\right)$
\STATE $i=i+1$
\ENDWHILE
\end{algorithmic}
\end{algorithm}
\subsection{Convergence and Computational Complexity}
The proposed beam squint-aware scheme of the equivalent fully digital precoder is based on fractional programming.
In particular, Dinkelbach's algorithm converts the original fractional problem into a series of subproblems, and each subproblem can be handled iteratively through the Lagrangian dual transform and the quadratic transform.
Note that the numerator and denominator of the objective function EE are both continuous and positive-valued functions of the variable $\cB$.
Subsequently, based on \cite[Theorem 2.2]{rodenas1999extensions}, the termination of Dinkelbach's algorithm can be achieved in a finite number of iterations.
In each iteration of Dinkelbach's algorithm, the objective of the subproblem is monotonically nondecreasing.
Then, the subproblem is guaranteed to converge to a stationary point of the objective function \cite[Proposition 2]{shen2018fractional}.
Therefore, the convergence of \textbf{Algorithm \ref{alg:algdasam}} is  confirmed.

The computation complexity for \textbf{Algorithm \ref{alg:algdasam}} mainly lies in the following aspects.
First, we assume that Dinkelbach's algorithm terminates in $I$ iterations \cite{rodenas1999extensions} and focus on the computational complexity for each subproblem.
In particular, both updates of $\lambda_{k,m}$ and $\rho_{k,m}$ require the complexity of $\mathcal{O}(N_{\rm t})$.
The complexity to compute the precoding vector $\bb_{k,m}$ depends on the pseudo-inverse operation, which is given by $\mathcal{O}(N_{\rm t}^3)$.
Note that the bisection search process in the computation of $\bb_{k,m}$ contains only a small number of iterations \cite{shi2011iteratively} and thus, can be ignored.
In conclusion, assuming that in average, the convergence of each subproblem involves $J$ iterations, the whole computational complexity of \textbf{Algorithm \ref{alg:algdasam}} is expressed as $\mathcal{O}\left(IJKM\left(N_{\rm t}^3+2N_{\rm t}\right)\right)$.
\section{Beam Squint-Aware Hybrid Precoding}\label{sec:JCS_hp}
We denote the equivalent fully digital precoder for the communications-only scenario as $\bB_{\rm com}[m]$ for the $m$th subcarrier. Then, we aim at minimizing the Euclidean distance between the product of the digital/analog precoders and the equivalent fully digital as well as sensing precoders.
Specifically, we formulate a weighted sum minimization problem to design the digital and the analog precoders at the $m$th subcarrier, expressed as \cite{zi2016energy,kaushik2021hardware}
    \begin{subequations}\label{eq:mind}
    \begin{align}
    \mathcal{Q}_1^m: \mathop{\mathrm{minimize}}\limits_{\substack{\bW_{\rm RF}[m],\\ \bW_{\rm BB}[m],\\ \bU[m]}}&\ \ f\left(\bW_{\rm RF}[m],\bW_{\rm BB}[m],\bU[m]\right)\label{eq:minda}\\
    {\rm s.t.}\ \ \ \ \ & \ \ \bW_{\rm RF}[m]\in \mathcal{S},\ \forall m,\label{eq:mindb}\\
    & \ \ ||\mathbf{B}_{\rm com}[m]||_F^2=||\bW_{\rm RF}[m]\bW_{\rm BB}[m]||_F^2,\ \forall m,\label{eq:mindc}\\
    &\ \ \bU[m]\bU^H[m]=\bI_{P_{\rm r}},\ \forall m, \label{eq:mindd}
    \end{align}
    \end{subequations}
where the expression for the objective function is shown in Eq. \eqref{eq:obj4} on the top of the next page, $\mathcal{S}\in \{\mathcal{S}_{\rm FC}, \mathcal{S}_{\rm PC}\}$ and the definitions of $\mathcal{S}_{\rm FC}, \mathcal{S}_{\rm PC}$ are presented in \eqref{eq:vconstr}.
The weighting coefficient $\zeta\in [0,1]$ can be set according to the different system requirements \cite{kaushik2021hardware}.
\newcounter{TempEqCnt4} 
\setcounter{TempEqCnt4}{\value{equation}} 
\setcounter{equation}{48} 
\begin{figure*}[ht] 
    \begin{align}\label{eq:obj4}
    f\left(\bW_{\rm RF}[m],\bW_{\rm BB}[m],\bU[m]\right)=\zeta\abs{\abs{\bW_{\rm RF}[m]\bW_{\rm BB}[m]-\bB_{\rm com}[m]}}_F^2+(1-\zeta)\abs{\abs{\bW_{\rm RF}[m]\bW_{\rm BB}[m]-\bB_{\rm ss}[m]\bU[m]}}_F^2
    \end{align}
    \hrule
\end{figure*}

Problem $\mathcal{Q}_1^m$ is not easy to handle due to the non-convexity in both the objective function and the constraints.
In addition, every non-zero element in the RF analog precoder $\bW_{\rm RF}[m]$ is of unit-modulus, making the problem harder.
Then, an alternating optimization framework is adopted to iteratively design the hybrid precoders as well as the unitary matrices. In particular, the methods for both the fully and partially connected structures are developed in the following, which can be applied for arbitrary $m$ and thus, the index $m$ is omitted for brevity.
\subsection{Fully Connected Structure}
For the fully connected structure, the RF analog precoder satisfies the condition $\bW_{\rm RF}\in \mathcal{S}_{\rm FC}$.
In the following, we alternatingly optimize the three variables $\bW_{\rm RF},\bW_{\rm BB},\bU$ by sequentially updating one of them with the other two fixed.

\subsubsection{Update of the Unitary Matrix $\bU$}
First, given $\bW_{\rm RF},\bW_{\rm BB}$, the original problem in \eqref{eq:mind} can be converted into an orthogonal Procrustes problem \cite{liu2019hybrid}, given by
    \begin{subequations}\label{eq:oppu}
    \begin{align}
    \mathcal{Q}_2: \mathop{\mathrm{minimize}}\limits_{\bU}&\ \ \abs{\abs{\bW_{\rm RF}\bW_{\rm BB}-\bB_{\rm ss}\bU}}_F^2\label{eq:oppua}\\
    {\rm s.t.} &\ \ \bU\bU^H=\bI_{P_{\rm r}}, \label{eq:oppub}
    \end{align}
    \end{subequations}
which admits an analytical solution by using the singular value decomposition (SVD). The update of the unitary matrix $\bU$ is given by
\begin{align}\label{eq:optU}
\bU=\bQ\bI_{P_{\rm r}\times K}\bR,
\end{align}
where $\bQ$ and $\bR$ can be obtained from the SVD of $\bB_{\rm ss}^H\bW_{\rm RF}\bW_{\rm BB}$ as $\bQ\mathbf{\Sigma}\bR=\bB_{\rm ss}^H\bW_{\rm RF}\bW_{\rm BB}$, and $\bI_{P_{\rm r}\times K}=[\bI_{P_{\rm r}},\mathbf{0}]$.
\subsubsection{Update of the Digital Precoder $\bW_{\rm BB}$}
With fixed $\bW_{\rm RF}$ and $\bU$, problem $\mathcal{Q}_1$ is rewritten as
    \begin{subequations}
    \begin{align}\label{eq:optw1}
    \mathcal{Q}_3: \mathop{\mathrm{minimize}}\limits_{\bW_{\rm BB}}&\ \ \zeta\abs{\abs{\bW_{\rm RF}\bW_{\rm BB}-\bB_{\rm com}}}_F^2\notag\\
    &\quad +(1-\zeta)\abs{\abs{\bW_{\rm RF}\bW_{\rm BB}-\bB_{\rm ss}\bU}}_F^2\\
    {\rm s.t.} &\ \ \abs{\abs{\bB_{\rm com}}}_F^2=\abs{\abs{\bW_{\rm RF}\bW_{\rm BB}}}_F^2.
    \end{align}
    \end{subequations}
Since the corresponding objective function is the sum of two Frobenius norms, problem $\cQ_3$ can be simplified as
    \begin{subequations}
    \begin{align}
    \mathcal{Q}_4: \mathop{\mathrm{minimize}}\limits_{\bW_{\rm BB}}&\ \ \abs{\abs{\bA\bW_{\rm BB}-\bC}}_F^2 \label{eq:optw2a}\\
    {\rm s.t.} &\ \ \abs{\abs{\bB_{\rm com}}}_F^2=\abs{\abs{\bA\bW_{\rm BB}}}_F^2,\label{eq:optw2b}
    \end{align}
    \end{subequations}
where $\bA=\left[\sqrt{\zeta}\bW_{\rm RF}^T,\sqrt{1-\zeta}\bW_{\rm RF}^T\right]^T\in \bbC^{2N_{\rm t}\times M_{\rm t}}$, $\bC=\left[\sqrt{\zeta}\bB_{\rm com}^T,\sqrt{1-\zeta}\bU^T\bB_{\rm ss}^T\right]^T\in \bbC^{2N_{\rm t}\times K}$.
Note that the auxiliary matrix $\bA$ satisfies $\bA^H\bA=\bW_{\rm RF}^H\bW_{\rm RF}$, which can be employed to simplify the computational complexity in the following.
To handle this problem, we first ignore the constraint in \eqref{eq:optw2b}.
Then, the relaxed problem is a least-squares one and has a closed-form solution, which is given by \cite{boyd2004convex}
\begin{align}\label{eq:optW}
\bW_{\rm BB}=\left(\bA^H\bA\right)^{-1}\bA^H\bC=\left(\bW_{\rm RF}^H\bW_{\rm RF}\right)^{-1}\bA^H\bC.
\end{align}
After that, we normalize the digital precoder $\bW_{\rm BB}$ as follows \cite{yu2016alternating,xiao2018joint}
    \begin{equation}
    \begin{aligned}
    \bW_{\rm BB}=\frac{||\bB_{\rm com}||_F}{||\bW_{\rm RF}\bW_{\rm BB}||_F}\bW_{\rm BB},
    \end{aligned}
    \end{equation}
and then the Euclidean distance in \eqref{eq:optw2a} can be sufficiently small \cite{yu2016alternating}.

\subsubsection{Update of the RF Analog Precoder $\bW_{\rm RF}$}
When the other two variables are fixed, the subproblem with respect to the RF analog precoder $\bW_{\rm RF}$ is given by
    \begin{subequations}\label{eq:optv1}
    \begin{align}
    \mathcal{Q}_5: \mathop{\mathrm{minimize}}\limits_{\bW_{\rm RF}}&\ \ \zeta\abs{\abs{\bW_{\rm RF}\bW_{\rm BB}-\bB_{\rm com}}}_F^2\notag\\
    &\quad+(1-\zeta)\abs{\abs{\bW_{\rm RF}\bW_{\rm BB}-\bB_{\rm ss}\bU}}_F^2\label{eq:optv1a}\\
    {\rm s.t.} & \ \ \bW_{\rm RF}\in \mathcal{S}_{\rm FC}. \label{eq:optv1b}
    \end{align}
    \end{subequations}
To simplify the above problem, we denote $\bG=\left[\sqrt{\zeta}\bW_{\rm BB},\sqrt{1-\zeta}\bW_{\rm BB}\right]\in \bbC^{M_{\rm t}\times 2K}$ and $\bT=\left[\sqrt{\zeta}\bB_{\rm com},\sqrt{1-\zeta}\bB_{\rm ss}\bU\right]\in \bbC^{N_{\rm t}\times 2K}$, and thus, problem $\mathcal{Q}_5$ can be rewritten as
    \begin{subequations}\label{eq:optv2}
    \begin{align}
    \mathcal{Q}_6: \mathop{\mathrm{minimize}}\limits_{\bW_{\rm RF}}&\ \ \abs{\abs{\bW_{\rm RF}\bG-\bT}}_F^2\label{eq:optv2a}\\
    {\rm s.t.} & \ \ \bW_{\rm RF}\in \mathcal{S}_{\rm FC},\label{eq:optv2b}
    \end{align}
    \end{subequations}
which can be tackled with a majorization-minimization (MM)-based method \cite{arora2019hybrid}.
In particular, let $\bY=\bG\bG^H$ and denote its maximum eigenvalue as $\lambda_{\rm max}(\bY)$. 
Then, the RF analog precoder $\bW_{\rm RF}$ can be updated as \cite{arora2019hybrid}
\begin{align}\label{eq:optV}
\bW_{\rm RF}=\exp\left\{-\jmath\angle \bZ^T\right\},
\end{align}
where $\bZ=\bG\bT^H-(\bY-\lambda_{\rm max}(\bY)\bI_{M_{\rm t}})\bW_{\rm RF}^H$.
The procedure for the design of the hybrid precoder is summarized in \textbf{Algorithm \ref{alg:algtrhpf}}.

\begin{algorithm}[!t]
\caption{Hybrid Precoding with the Fully Connected Structure}
\label{alg:algtrhpf}
\begin{algorithmic}[1]
\REQUIRE Equivalent fully digital communication precoding vector $\bB_{\rm com}$, threshold $\epsilon_2$, the number of maximum iterations $c_{\rm max}$
\STATE Initialize $\bW_{\rm RF}^{(0)}$, $\bW_{\rm BB}^{(0)}$, and $\bU^{(0)}$
\FOR{$m=1:M$}
\REPEAT
\STATE
Calculate the value of the objective function $f^{(c)}(\bW_{\rm RF},\bW_{\rm BB},\bU)$
\STATE
Update $\bU$, $\bW_{\rm BB}$, and $\bW_{\rm RF}$, based on Eqs. \eqref{eq:optU}, \eqref{eq:optW}, and \eqref{eq:optV}
\STATE $c = c+1$
\UNTIL $c\geq c_{\rm max}$ or $\abs{f^{(c)}-f^{(c-1)}}<\epsilon_2$
\ENDFOR
\STATE The digital precoder is normalized as $\displaystyle \mathbf{W}_{\rm BB}=\frac{||\mathbf{B}_{\rm com}||_F}{||\bW_{\rm RF}\bW_{\rm BB}||_F}\mathbf{W}_{\rm BB}$
\end{algorithmic}
\end{algorithm}
\subsection{Partially Connected Structure}
The implementation of the RF analog precoder with the partially connected structure differs from the one with the fully connected counterpart, which is constrained by $\bW_{\rm RF}\in \cS_{\rm PC}$,  and can be expressed as $\left[\bW_{\rm RF}\right]_{i,j}=\exp\left\{\jmath \phi_{i,j}\right\}, \ \forall i, \ j=\left\lceil i/N_{\rm g}\right\rceil$.
Note that for the partially connected structure, the column index of each element of the matrix $\bW_{\rm RF}$ is dependent on the row index.
Moreover,
due to the block diagonal structure of $\bW_{\rm RF}$, the equality constraint in \eqref{eq:mindc} can be simplified as $||\bW_{\rm RF}\bW_{\rm BB}||_F^2=N_{\rm g}||\bW_{\rm BB}||_F^2=||\bB_{\rm com}||_F^2$.
\subsubsection{Update of the Unitary Matrix $\bU$}
The algorithm for this subproblem is exactly the same as the one for the fully connected case, as shown in \eqref{eq:optU}.
\subsubsection{Update of the Digital Precoder $\bW_{\rm BB}$}
For the partially connected case, with fixed $\bW_{\rm RF}$ and $\bU$, this subproblem can be expressed as
    \begin{subequations}
    \begin{align}\label{eq:optwpc}
    \mathcal{Q}_7: \mathop{\mathrm{minimize}}\limits_{\bW_{\rm BB}}&\ \ \abs{\abs{\bA^H\bC-\bW_{\rm BB}}}_F^2\\
    {\rm s.t.} &\ \ ||\bW_{\rm BB}||_F=\frac{||\bB_{\rm com}||_F}{\sqrt{N_{\rm g}}},\label{eq:optwpcb}
    \end{align}
    \end{subequations}
where matrices $\bA$ and $\bC$ follow the same definitions as the fully connected case. Note that $\cQ_7$ is a projection problem which admits a closed-form solution given by
\begin{align}\label{eq:normwpc}
\mathbf{W}_{\rm BB}=\frac{||\mathbf{B}_{\rm com}||_F}{\sqrt{N_{\rm g}}}\frac{\bA^H\bC}{||\bA^H\bC||_F}.
\end{align}
\subsubsection{Update of the RF Analog Precoder $\bW_{\rm RF}$}
With $\bW_{\rm BB}$ and $\bU$ fixed, the subproblem for determining the RF analog precoder $\bW_{\rm RF}$ is shown as \cite{kaushik2021hardware}
\begin{align}
\mathcal{Q}_8: \mathop{\mathrm{minimize}}\limits_{\bW_{\rm RF}}&\ \ \abs{\abs{\ba-\exp\left\{\jmath\phi_{i,j}\right\}\bp}}_2^2\\
    {\rm s.t.} &\ \ \phi_{i,j}\in (0,2\pi],
\end{align}
where $\ba=\left[\sqrt{\zeta}[\bB_{\rm com}]_{i,:},\sqrt{1-\zeta}[\bB_{\rm ss}\bU]_{i,:}\right]$ and $\bp=\left[\sqrt{\zeta}[\bW_{\rm BB}]_{j,:},\sqrt{1-\zeta}[\bW_{\rm BB}]_{j,:}\right]$. Then, the update of the RF analog precoder is given by
\begin{align}\label{eq:udpv}
[\bW_{\rm RF}]_{i,j}=\exp\{\jmath \angle \left(\ba\bp^H\right)\},\ \forall i, \ \forall j=\left\lceil \frac{i}{N_{\rm g}}\right\rceil.
\end{align}
The procedure for designing the hybrid precoders with the partially connected structure is sketched in \textbf{Algorithm \ref{alg:algtrhpp}}.
\begin{algorithm}[htbp]
\caption{Hybrid Precoding with the Partially Connected Structure}
\label{alg:algtrhpp}
\begin{algorithmic}[1]
\REQUIRE Equivalent fully digital communication precoding vector $\bB_{\rm com}$, threshold $\epsilon_3$, the number of maximum iterations $c_{\rm max}$
\STATE Initialize $\bW_{\rm RF}^{(0)}$, $\bW_{\rm BB}^{(0)}$, and $\bU^{(0)}$
\FOR{$m=1:M$}
\REPEAT
\STATE
Calculate the value of the objective function $f^{(c)}(\bW_{\rm RF},\bW_{\rm BB},\bU)$
\STATE
Update $\bU$, $\bW_{\rm BB}$, and $\bW_{\rm RF}$, based on Eqs. \eqref{eq:optU}, \eqref{eq:normwpc}, and \eqref{eq:udpv}
\STATE $c = c+1$
\UNTIL $c\geq c_{\rm max}$ or $\abs{f^{(c)}-f^{(c-1)}}<\epsilon_3$
\ENDFOR
\end{algorithmic}
\end{algorithm}
\subsection{Convergence and Computational Complexity}
Both \textbf{Algorithms \ref{alg:algtrhpf}} and \textbf{\ref{alg:algtrhpp}} can be regarded as block coordinate descent methods, which are guaranteed to converge \cite{beck2015convergence}.
In addition, for the fully connected case, the major computational complexity for each iteration lies in the pseudo-inverse operation of the subproblem to update the baseband digital precoder $\bW_{\rm BB}$, which can be estimated as $\mathcal{O}\left(N_{\rm t}^3\right)$.
Besides, the algorithm for the partially connected structure mainly involves the SVD process and the calculation of Eqs. \eqref{eq:normwpc}, \eqref{eq:udpv}, whose computational complexity are given by $\mathcal{O}\left(P_{\rm r}K^2\right)$, $\mathcal{O}\left(M_{\rm t}N_{\rm t}K\right)$, and $\mathcal{O}\left(N_{\rm t}KM_{\rm t}\right)$, respectively.
Let $I$ and $J$ denote the average iterations to design the hybrid precoders with the fully connected and partially connected structure for each subcarrier.
Then, the computational complexity for \textbf{Algorithms \ref{alg:algtrhpf}} and \textbf{\ref{alg:algtrhpp}} are given by $\mathcal{O}\left(MIN_{\rm t}^3\right)$ and $\mathcal{O}\left(MJ\left(P_{\rm r}K^2+2M_{\rm t}N_{\rm t}K\right)\right)$, respectively.
\section{Simulations}\label{sec:sim}
This section evaluates the performance of the proposed LEO satellite ISAC scheme.
The considered system operates in the Ka-band \cite{3gpp} and some typical simulation parameters are listed in \tabref{tab:test}.
At the transmitter, the antennas on the x- and y-axes of the UPA are assumed to be identical, i.e., $N_{\rm t}^{\rm x}=N_{\rm t}^{\rm y}$, each varying from 20 to 48.
Then, the total number of the UPA antennas is given by $N_{\rm t}=N_{\rm t}^{\rm x}\times N_{\rm t}^{\rm y}$, valued from 400 to 2304.
The downlink channel power can be calculated by \cite{li2020downlink}
\begin{align}\label{eq:normchp}
\gamma_k = G_{\rm sat}G_{\rm ut}N_\mathrm{t}\left(\frac{c}{4\pi f_{\rm c}d_0}\right)^2, \forall k,
\end{align}
where $G_{\rm sat}$ and $G_{\rm ut}$ denote the antenna gains of the transmitter and the receiver, respectively.
The noise power is defined as $N_0^{\rm c}=N_0^{\rm r}=k_{\rm B}\Delta_BT_{\rm n}$ with the Boltzmann constant $k_{\rm B}=1.38\times 10^{-23}$ J$\cdot\text{K}^{-1}$ and the noise temperature $T_{\rm n}=300$ K \cite{li2020downlink}.
\begin{table*}[!t]
\caption{Simulation Parameters}\label{tab:test}
\centering
\ra{1.3}
\footnotesize
\begin{tabular}{LLR}
\toprule
&\ Parameter &\ Value  \\
\midrule
\multirow{8}{*}{Channel} &\cellcolor{lightblue} System bandwidth $B_{\rm w}$ & \cellcolor{lightblue} Up to 800 MHz\\
&\ Carrier frequency $f_c$ &\ 20 GHz \\
&\cellcolor{lightblue} Speed of light $c$ &\cellcolor{lightblue} 3$\times 10^8$ m/s\\
&\ Carrier wavelength $\lambda_c$ &\ 0.015 m \\
&\cellcolor{lightblue} Signal duration $T_{\rm s}$ &\cellcolor{lightblue}  1.25 ns\\
&\ Number of OFDM subcarriers $M$ &\ 40 \\
&\cellcolor{lightblue} Rician factor $\kappa_k$ & \cellcolor{lightblue} 12 dB\\
&\ Antenna gain $G_{\rm sat}$, $G_{\rm ut}$ &\ 3 dB\\
\midrule
\multirow{8}{*}{Satellite} & \cellcolor{lightblue} Orbit altitude $d_0$ &\cellcolor{lightblue} 1000 km\\
&\ Number of antennas $N_{\rm t}$ & 400 $\sim$ 2304\\
&\cellcolor{lightblue} Antenna spacing $r_{\rm x}$, $r_{\rm y}$ &\cellcolor{lightblue} $\lambda_c/2$ \\
&\ Number of RF chains $M_{\rm t}$ &\ 16 \\
&\cellcolor{lightblue} Efficiency of power amplifier $1/\xi$ &\cellcolor{lightblue} 0.5 \\
&\ Per-RF chain power consumption $P_{\rm RFC}$ & \ 338mW \\
&\cellcolor{lightblue} Static power consumption of local oscillator $P_{\rm LO}$ &\cellcolor{lightblue} 5 mW\\
&\ Static power consumption of baseband precoder $P_{\rm BB}$&\ 200 mW\\
\midrule
\multirow{4}{*}{UTs/Targets} &\cellcolor{lightblue} Number of UTs $K$ &\cellcolor{lightblue} 16\\
&\ Number of targets $\ P_{\rm r}$ &\ 4\\
&\cellcolor{lightblue} Space angle distribution of UT &\cellcolor{lightblue} $\cU(-1,1)$\\
&\multirow{2}{*}{\ Space angle pair of targets} &\ $(-0.3,0.7)$ $(0.6,-0.2)$ \\
& &\ $(-0.5,-0.9)$ $(0.4,0.8)$ \\
\bottomrule
\end{tabular}
\end{table*}

\begin{figure}[hb]
\centering
\subfloat[Algorithm \ref{alg:algdasam}.]{\includegraphics[width=0.33\textwidth]{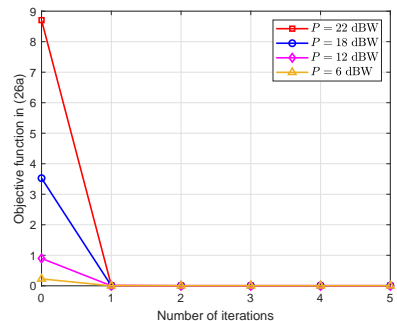}\label{fig_conver}}
\hfill
\subfloat[Algorithm \ref{alg:algtrhpf}.]{\includegraphics[width=0.33\textwidth]{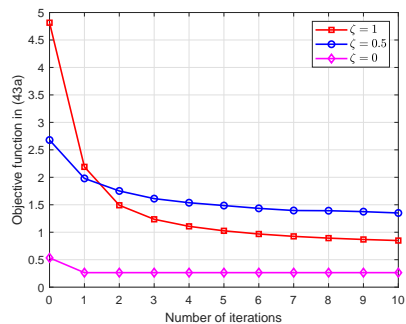}\label{fig_conver_fc}}
\hfill
\subfloat[Algorithm \ref{alg:algtrhpp}.]{\includegraphics[width=0.33\textwidth]{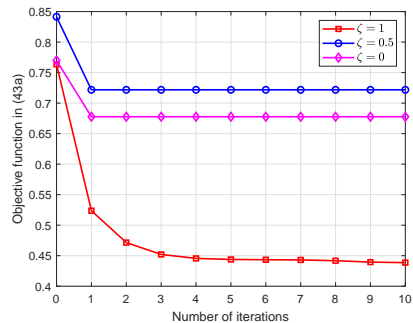}\label{fig_conver_pc}}
\caption{Convergence performance versus the number of iterations.}
\label{convergence}
\end{figure}

The average convergence performance for \textbf{Algorithms} \textbf{\ref{alg:algdasam}}, \textbf{\ref{alg:algtrhpf}}, and \textbf{\ref{alg:algtrhpp}} are shown in \figref{convergence}.
The corresponding objective functions are defined in Eqs. \eqref{eq:dbso} and \eqref{eq:minda}, respectively. In particular, \textbf{Algorithm \ref{alg:algdasam}} converges rapidly for different power budgets. Besides, there is a gradual trend regarding the convergence of \textbf{Algorithms \ref{alg:algtrhpf}} and \textbf{\ref{alg:algtrhpp}} with different values of weighting factors, which in general converges within a few iterations.

\begin{figure}[htbp]
\centering
\subfloat[Fully connected structure.]{\includegraphics[width=0.33\textwidth]{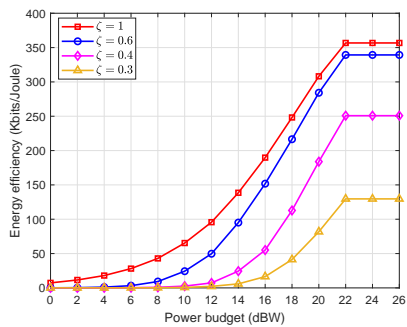}\label{EE_p_bs_zeta}}
\hfill
\subfloat[Partially connected structure.]{\includegraphics[width=0.33\textwidth]{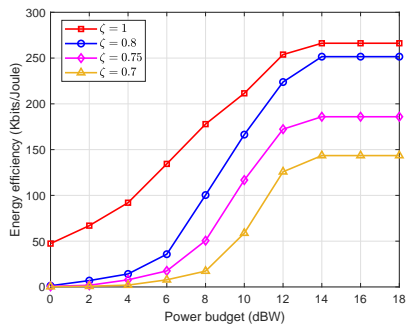}\label{EE_p_bs_dz_pc}}
\caption{EE performance versus power budget $P$ with $N_{\rm t}=576$ antennas under different weighting coefficients $\zeta$.}
\label{EE_p_fpbs_zeta}
\end{figure}

\figref{EE_p_fpbs_zeta} illustrates the EE performance for the communication module versus the power budget $P$ under different weighting coefficients $\zeta$. In both the fully and partially connected structures, for each $\zeta$, the value of EE follows the same upward trend, which can be categorized into three stages.
In particular, there is a slight rise in the value of EE before it bottoms out and grows rapidly.
Then, the growth rate declines and the EE value stays steady after its saturation point.
In fact, there exists an inherent saturation point for typical setups. After that point, an increase in the power consumption brings smaller gains in the data rate and thus, does not contribute to the performance of EE.
Besides, for the low or middle range transmit power scenarios, the deterministic factor of the EE performance is the static power consumption and thus, the partially connected structure, which consumes less power, performs better.
On the other hand, when the  transmit power is relatively high, the array gain dominates the EE performance and thus, the fully connected structure, which offers higher array gain, presents better performance.
Regarding the variation tendency with $\zeta$, better EE performance can be achieved with the increase of $\zeta$, since a larger weight is allocated to the communication module.
Note that when $\zeta$ is set to be one, the satellite ISAC systems reduce to communications-only ones.

\begin{figure*}[ht]
\centering
\subfloat[$\zeta=0.3$, $N_{\rm t}=576$.]{\includegraphics[width=0.33\textwidth]{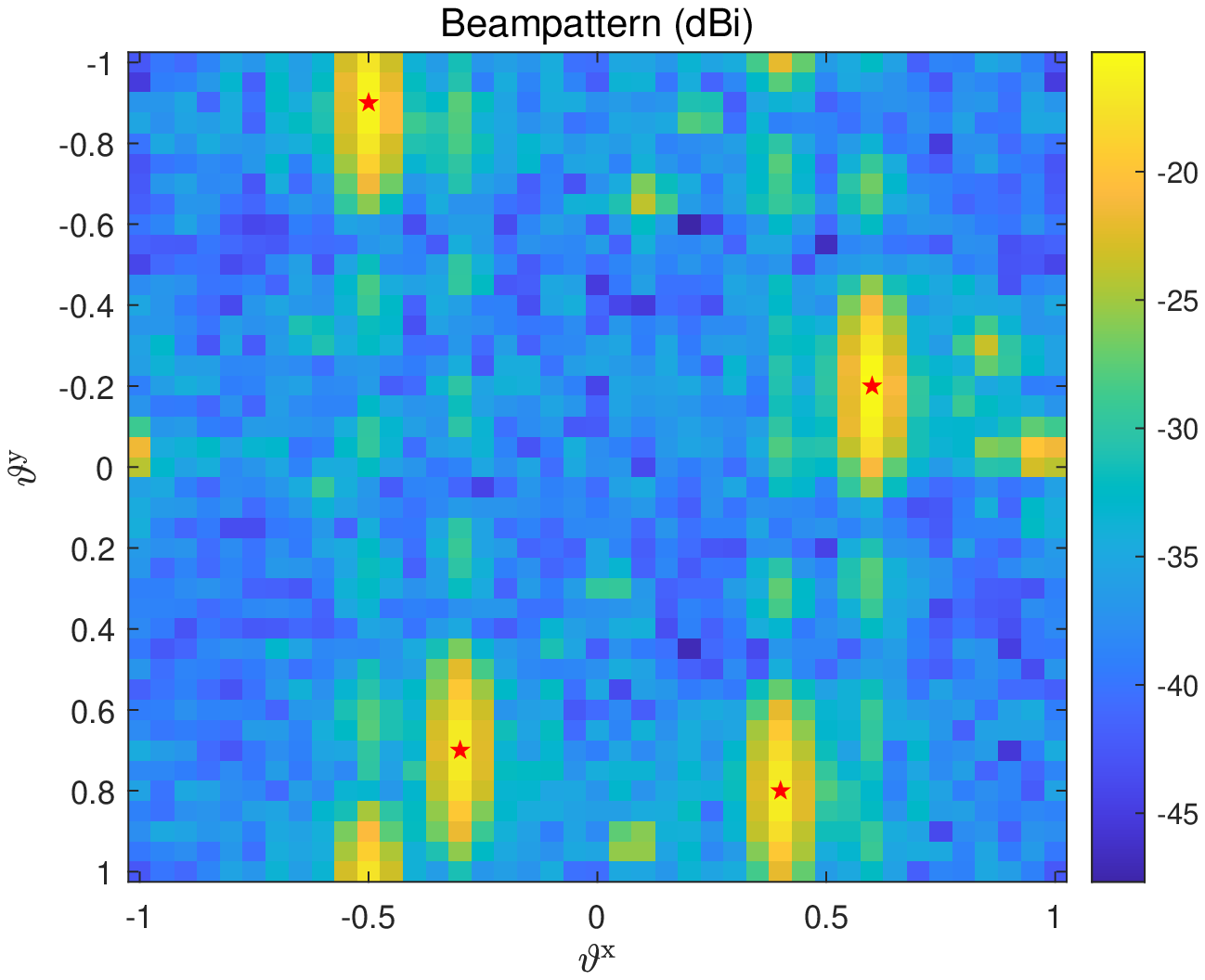}\label{beampattern_fc_bs_nt576_zeta03}}
\hfill
\subfloat[$\zeta=0.4$, $N_{\rm t}=576$ .]{\includegraphics[width=0.33\textwidth]{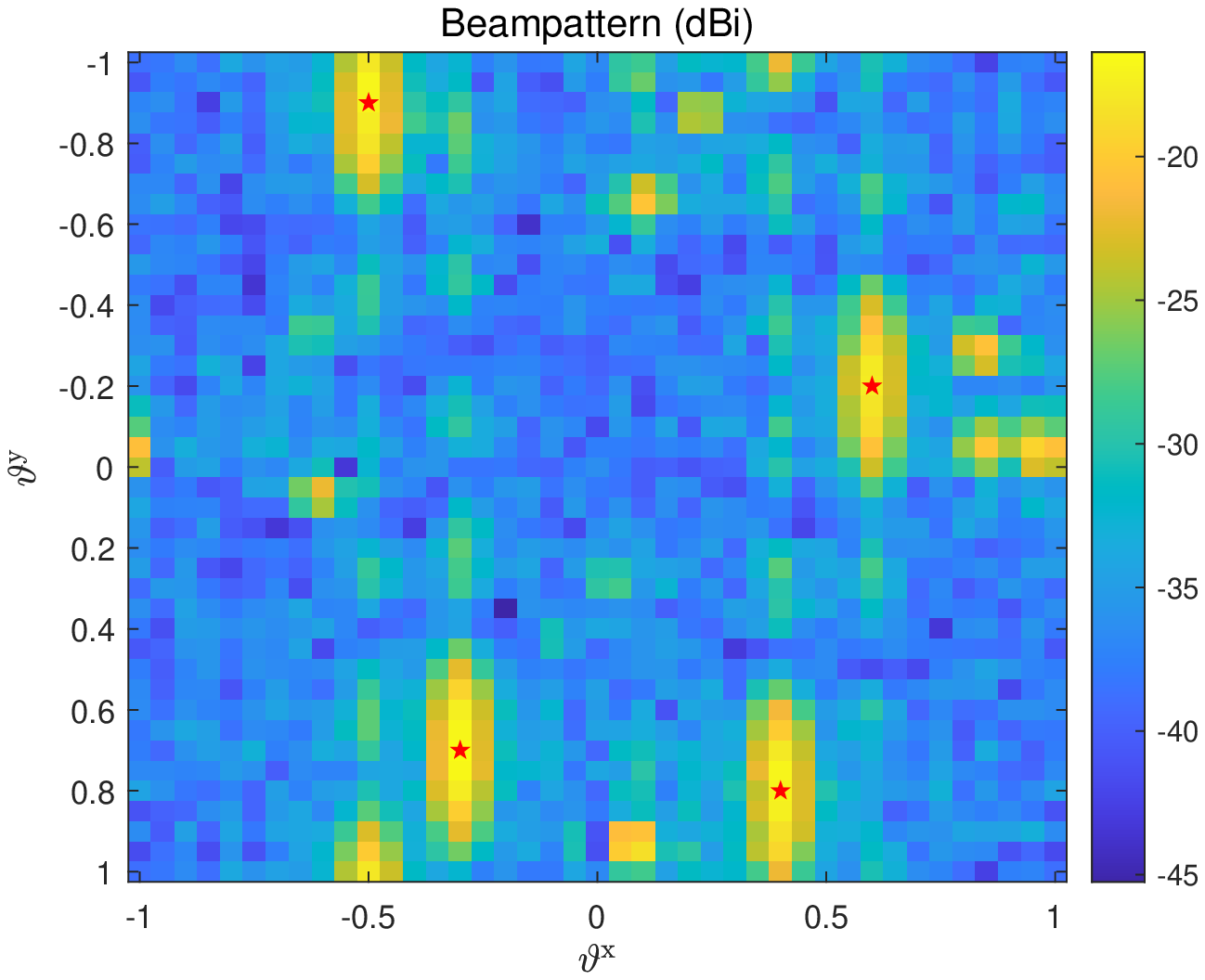}\label{beampattern_fc_bs_nt576_zeta04}}
\hfill
\subfloat[$\zeta=0.4$, $N_{\rm t}=2304$ .] {\includegraphics[width=0.33\textwidth]{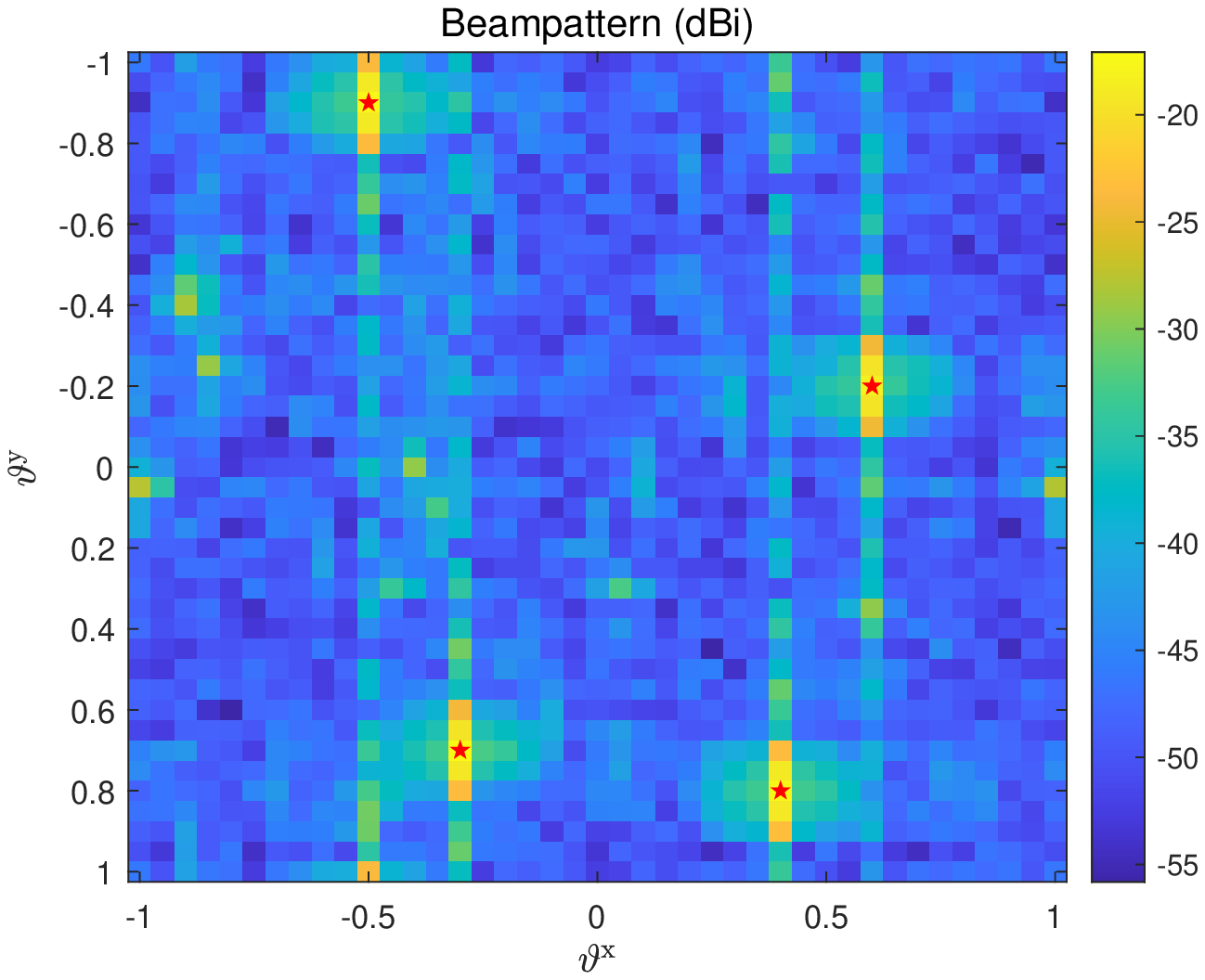}\label{beampattern_fc_bs_nt576_zeta04}}
\caption{Beampattern versus space angles with a fully connected hybrid precoder.}
\label{beampattern1}
\end{figure*}

\begin{figure*}[ht]
\centering
\subfloat[$\zeta=0.6$, $N_{\rm t}=576$.]{\includegraphics[width=0.33\textwidth]{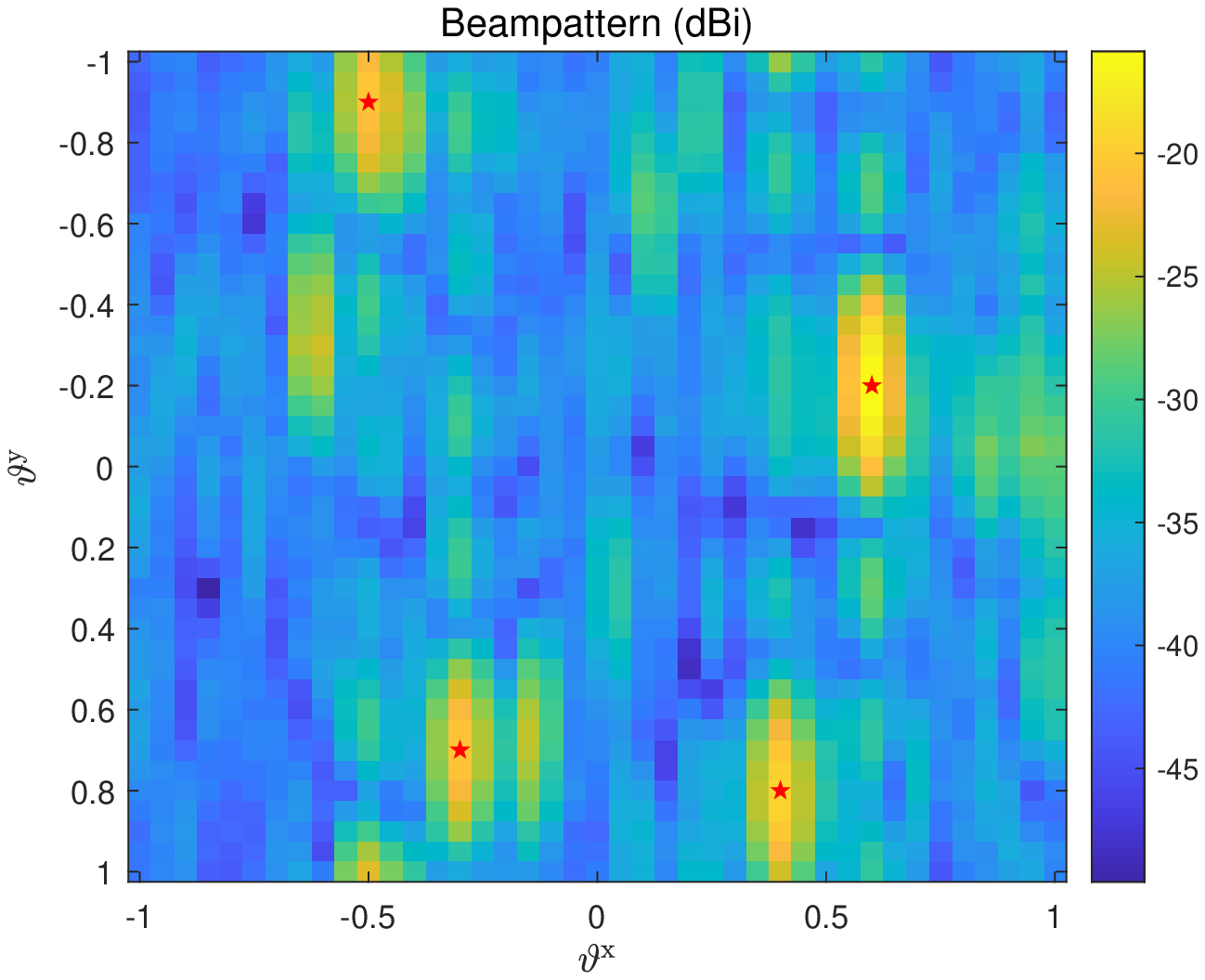}\label{beampattern_pc_bs_nt576_zeta06}}
\hfill
\subfloat[$\zeta=0.7$, $N_{\rm t}=576$.]{\includegraphics[width=0.33\textwidth]{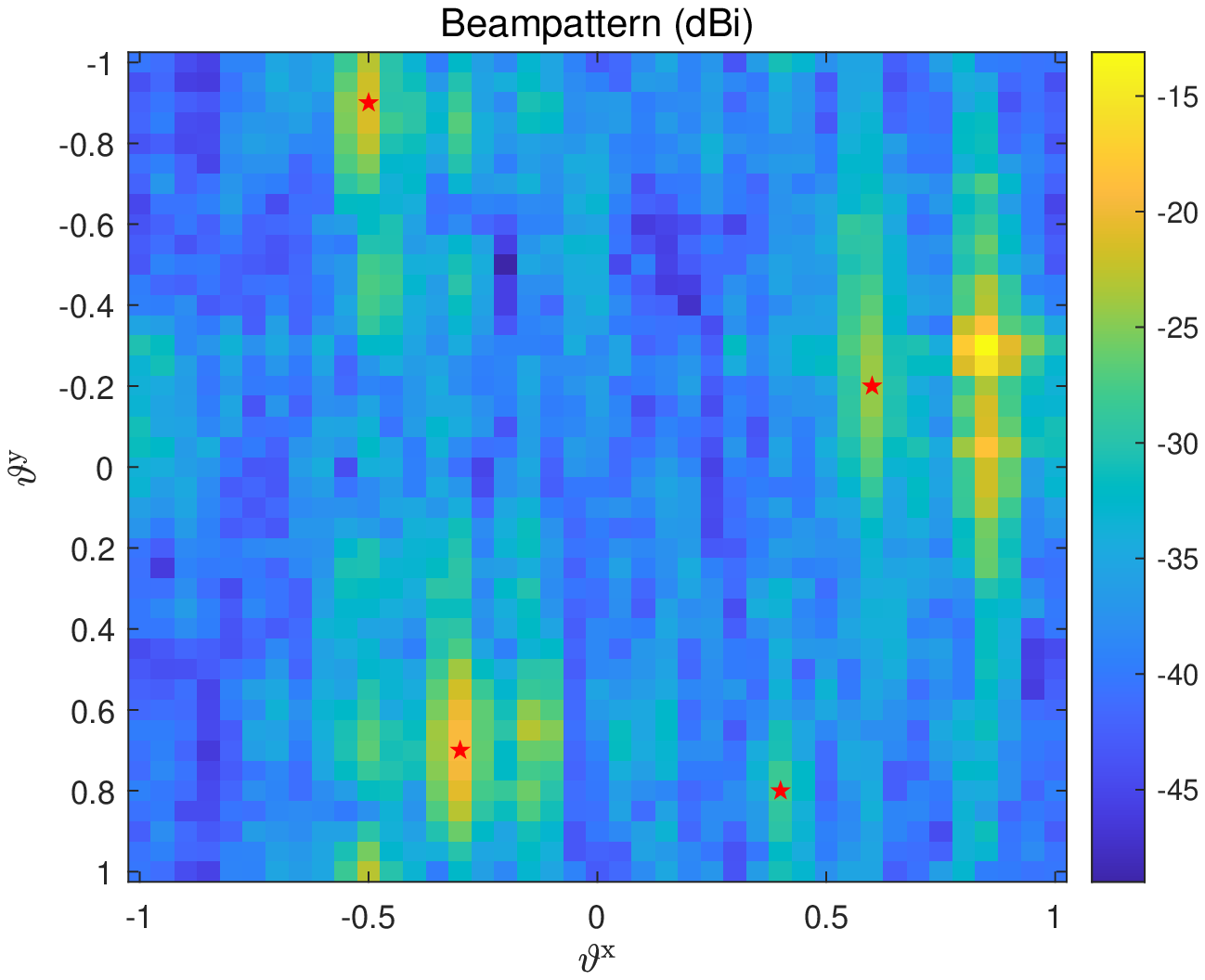}\label{beampattern_pc_bs_nt576_zeta07}}
\hfill
\subfloat[$\zeta=0.7$, $N_{\rm t}=2304$.] {\includegraphics[width=0.33\textwidth]{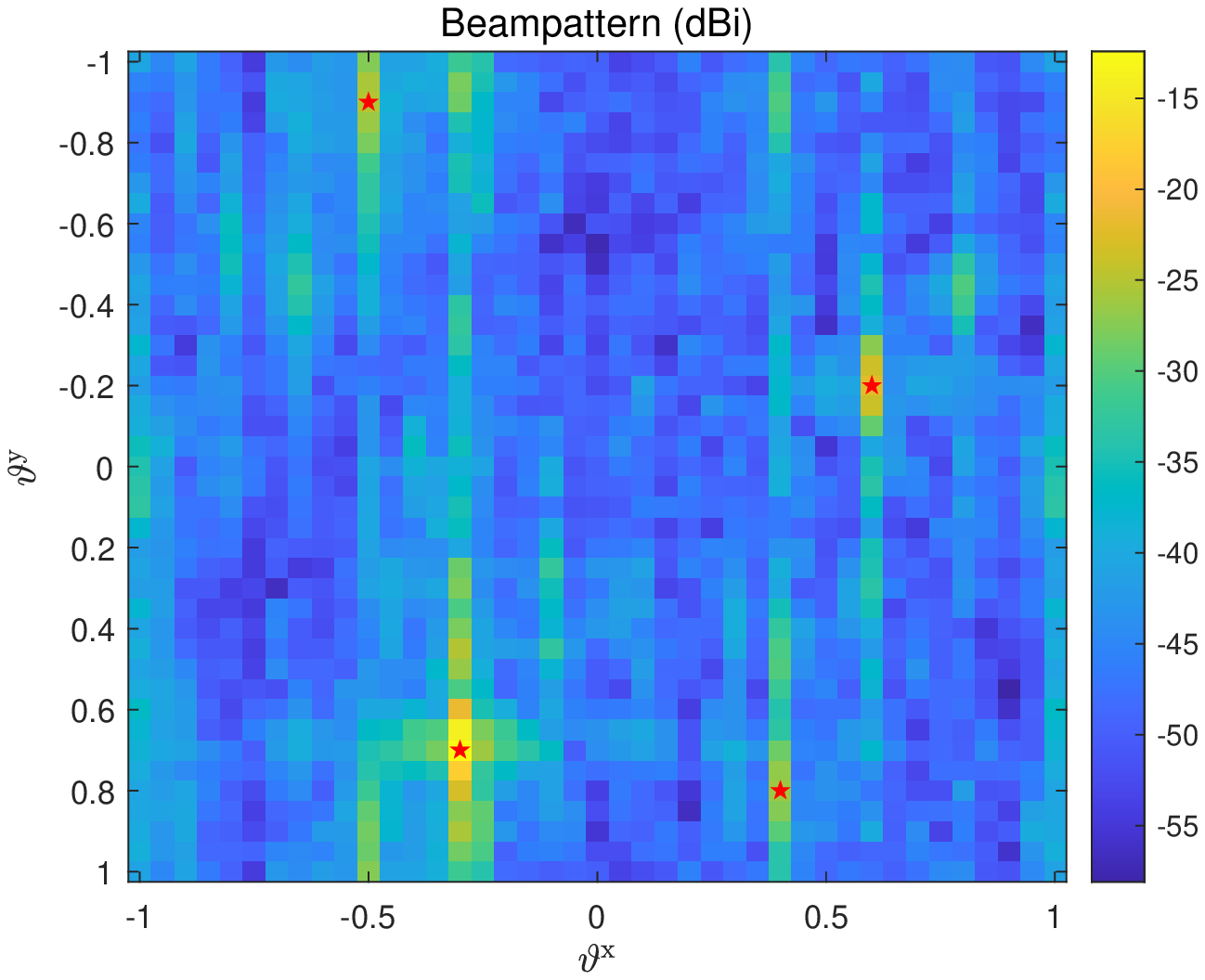}\label{beampattern_pc_bs_nt2304_zeta07}}
\caption{Beampattern versus space angles with a partially connected hybrid precoder.}
\label{beampattern2}
\end{figure*}

\begin{figure*}[ht]
\centering
\subfloat[Fully digital architecture.]{\includegraphics[width=0.33\textwidth]{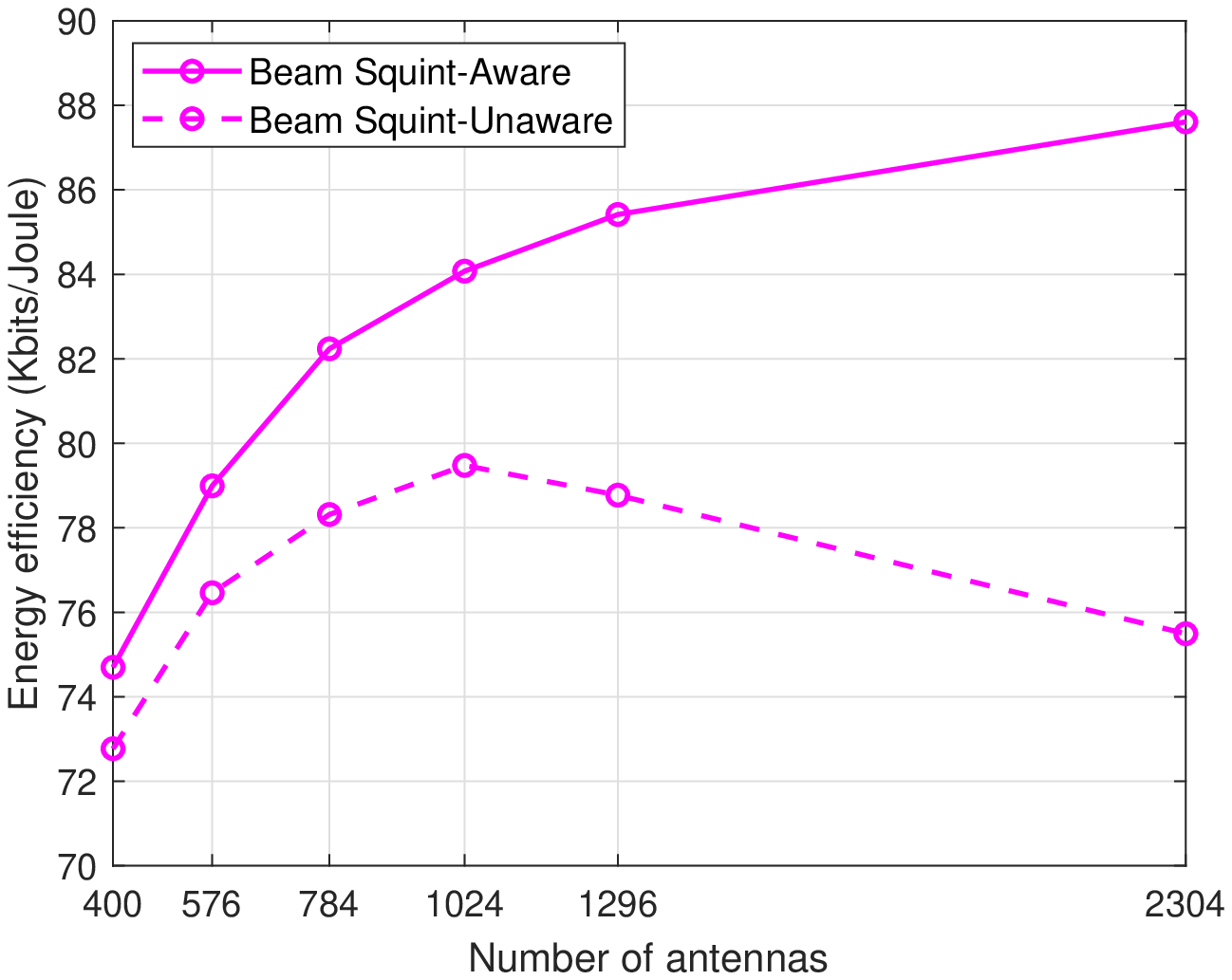}\label{fig_nt_fd_bs_p12}}
\hfill
\subfloat[Fully connected structure.]{\includegraphics[width=0.33\textwidth]{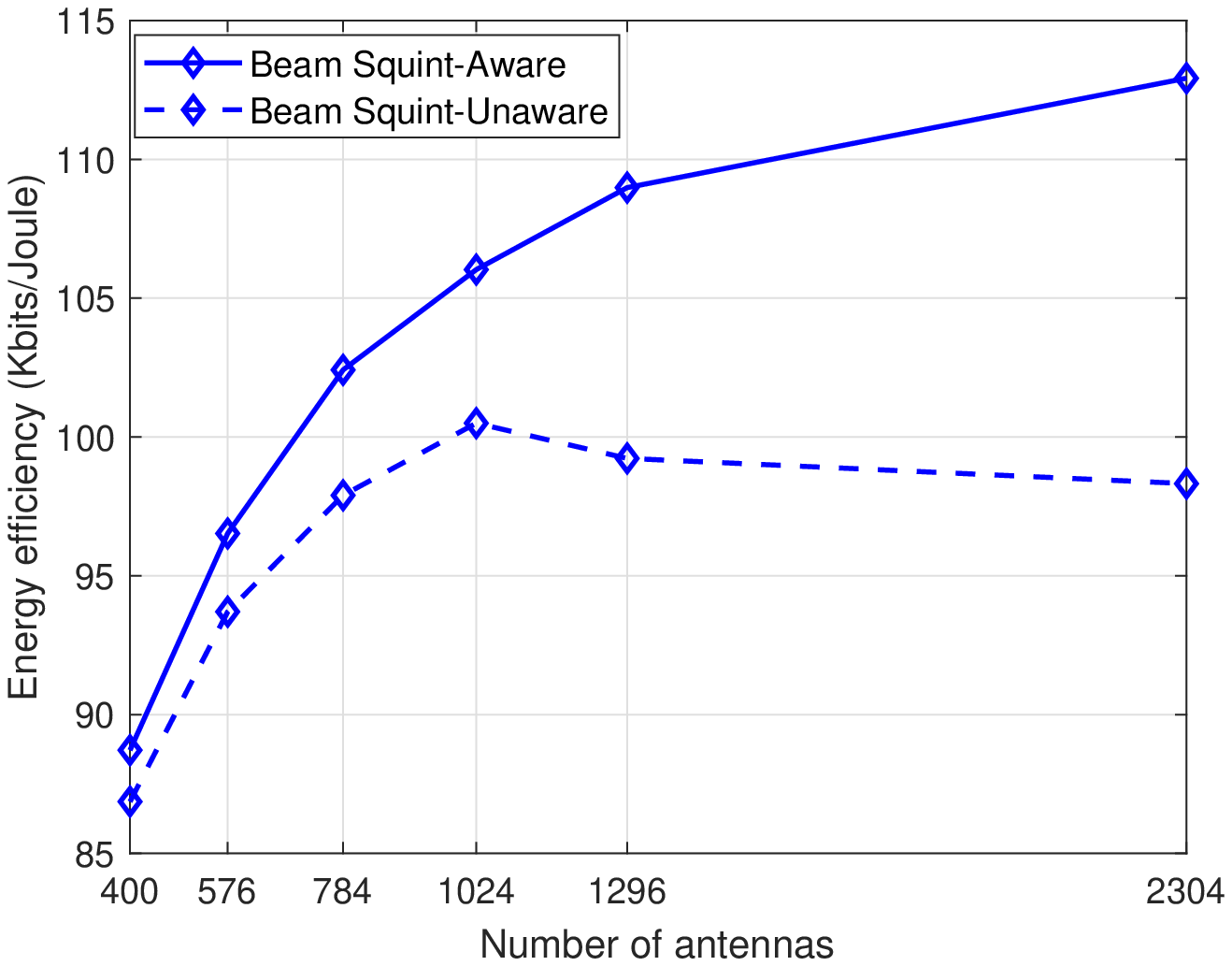}\label{fig_nt_fc_bs_p12}}
\hfill
\subfloat[Partially connected structure.]{\includegraphics[width=0.33\textwidth]{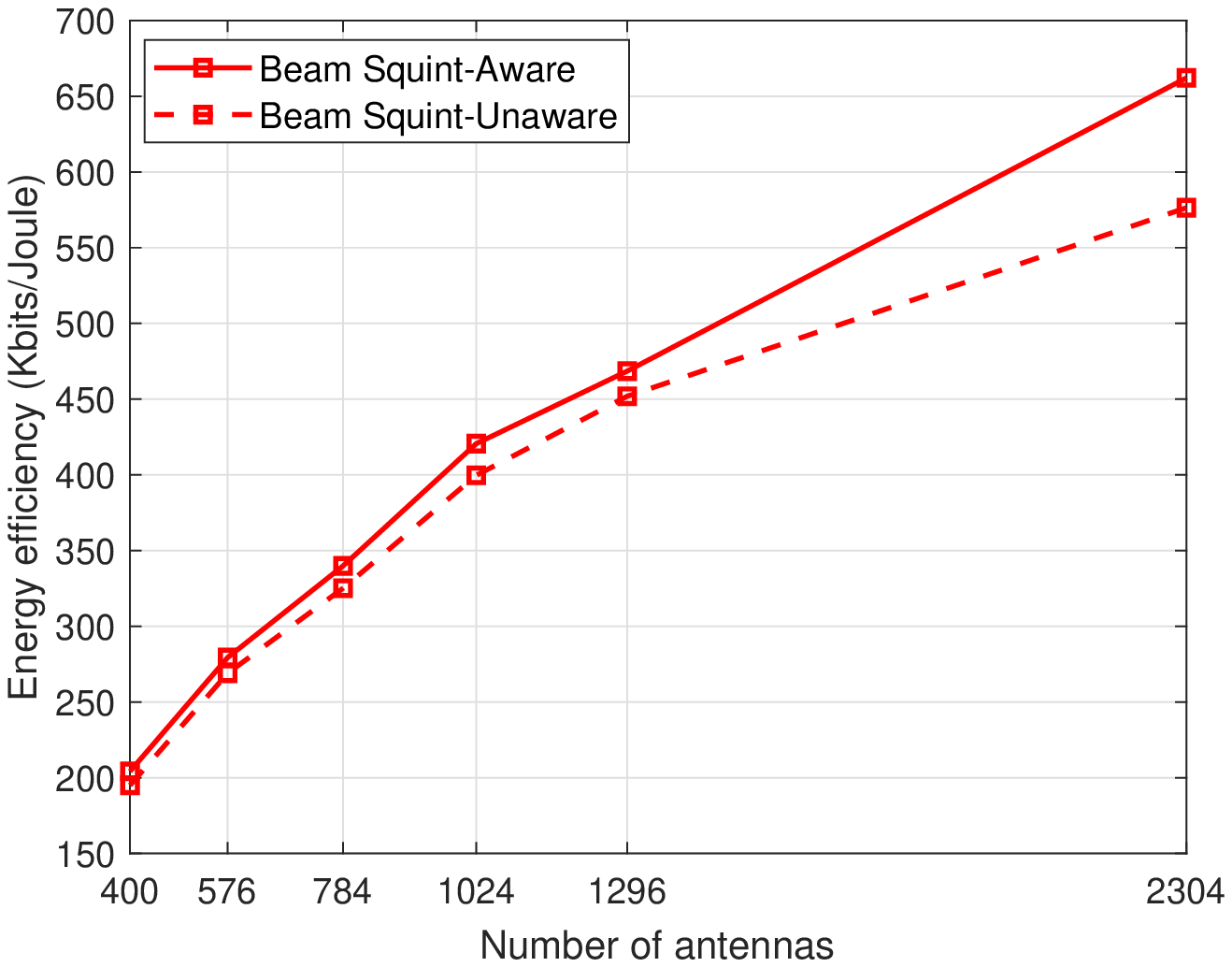}\label{fig_nt_pc_bs_p12_bw8}}
\caption{EE versus the number of antennas with $M_{\rm t}=K$ RF chains and transmission bandwidth $B_{\rm w}=800$ MHz under $P=12$ dBW for different architectures.}
\label{fig_bs_nt_arc}
\end{figure*}

Figs. \ref{beampattern1} and \ref{beampattern2} present the corresponding sensing beampattern with the consideration of the wide-band effects.
The red pentagrams locate the position of the targets.
With the increasing of the number of equipped antennas, the range resolution of sensing is improved and the disturbance of the communication module on the target sensing is significantly mitigated.
With the EE performance shown in \figref{EE_p_fpbs_zeta} as a reference, it can be observed that the beams can be efficiently steered towards the targets of interest while sustaining the performance of EE for the communication module.
In particular, for the fully connected structure, with a small weighting factor $\zeta=0.4$ assigned to the communication module, the EE performance sees a marked rise while the targets can be efficiently detected.
For the partially connected structure, with a weighting factor $\zeta=0.7$, both communications and sensing can be well performed.
Therefore, with different weighting factors, the adopted satellite ISAC systems can be flexibly adjusted to accommodate the different requirements of communications and target sensing.

\figref{fig_bs_nt_arc} evaluates the communication EE performance of the proposed beam squint-aware scheme with respect to the number of antennas. We adopt the beam squint-unaware scheme as the comparison baseline, which is tackled through \textbf{Algorithm} \textbf{\ref{alg:algdasam}} by setting $\bv_k(f_m)=\bv_k(0), \forall k,\ \forall m$.
In general, all the solid curves see a rising trend due to the increase of channel gain according to Eq. \eqref{eq:normchp}.
Besides, by further increasing the number of antennas, the considered system consumes more power and thus, the growth rates of the value of EE slow down.
Moreover, the proposed scheme with consideration of the beam squint effects outperforms that of the beam squint-unaware counterpart, especially for a system with a larger number of antennas.
In addition, with an increasing number of antennas, the time delay tends larger, and the impact of the beam squint deteriorates  the beam squint-unaware precoding method  even further.
Furthermore, the EE of the hybrid architecture outperforms the fully digital counterpart, especially with a large number of antennas.
This is due to the reason that for relatively small number of antennas, the static power consumed by the RF chains is not significantly high.
However, if the number of antennas is further increased, the static power consumption dominates the value of EE, and thus, the application of the hybrid architecture leads, in general, to a system with fewer RF chains and lower power consumption, resulting in improved EE performance.

   \begin{figure}[!t]
		\centering
		\includegraphics[width=0.33\textwidth]{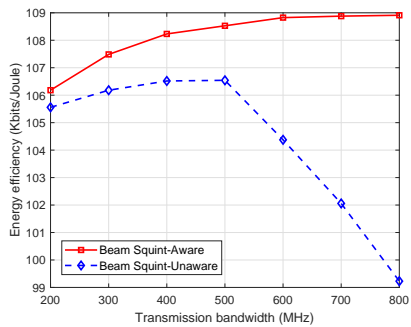}
		\caption{EE versus the transmission bandwidth with $M_{\rm t}=K$ RF chains and $N_{\rm t}=1296$ antennas under $P=12$ dBW.}
        \label{fig_bs_bw}
	\end{figure}

\begin{figure}[!t]
\centering
\subfloat[Beam squint-aware.]{\includegraphics[width=0.33\textwidth]{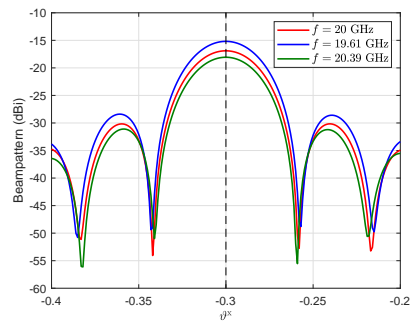}\label{beampattern_bs_h}}
\hfill
\subfloat[Beam squint-unaware.]{\includegraphics[width=0.33\textwidth]{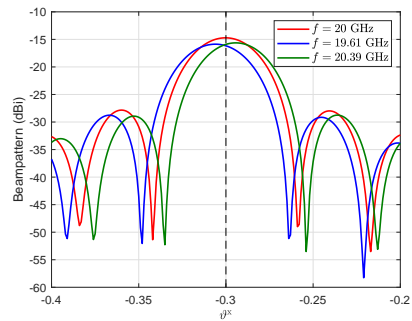}\label{beampattern_nbs_h}}
\hfill
\subfloat[Beam squint-aware.]{\includegraphics[width=0.33\textwidth]{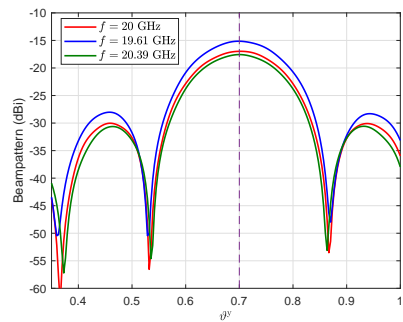}\label{beampattern_bs_v}}
\hfill
\subfloat[Beam squint-unaware.]{\includegraphics[width=0.33\textwidth]{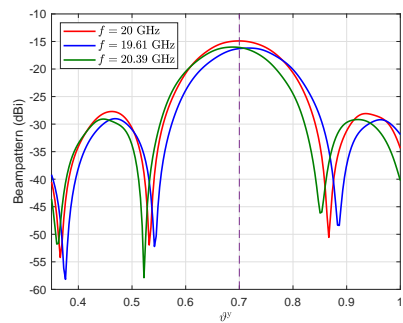}\label{beampattern_nbs_v}}
\caption{Beam squint example for sensing beampattern at the horizontal direction: (a), (b) and the vertical direction: (c), (d), with $N_{\rm t}=2304$ antennas and transmission bandwidth $B_{\rm w}=800$ MHz.}
\label{fig_bs_radar}
\end{figure}

\figref{fig_bs_bw} compares the communication EE performance of the proposed beam squint-aware precoding scheme with the beam squint-unaware one.
Note that the EE performance of the proposed beam squint-aware scheme sees a gradual growth with the increase of the bandwidth.
Furthermore, due to the beam squint effects, the performance gap between the two schemes becomes larger when the transmission bandwidth is improved.
The explanation for this trend initiates from the fact that the growth of the bandwidth leads to the reduction of the signal duration.
Hence, the ratio of the time delay over the UPA array and the signal duration increases. Consequently, a more negative influence is introduced on the beam squint-unaware scheme.

\figref{fig_bs_radar} depicts the sensing beampattern for the LEO satellite ISAC system for a target with the AoD $(-0.3,0.7)$.
With the adopted hybrid beam squint-aware scheme, the beam squint effects at both horizontal and vertical directions can be significantly mitigated, compared with the one without the consideration of the beam squint effects.
Besides, due to the utilization of the hybrid analog/digital architecture, there exists a small loss in the gain of the beam pattern at some subcarrier frequencies.

   \begin{figure}[htbp]
		\centering
		\includegraphics[width=0.33\textwidth]{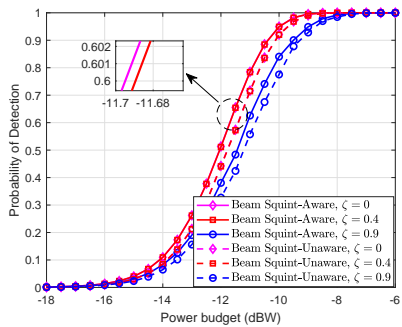}
		\caption{Detection probability $P_{\rm D}$ versus power budget $P$ with $N_{\rm t}=2304$ antennas and false-alarm probability $P_{\rm FA}=10^{-7}$.}
        \label{fig_pod_fc}
	\end{figure}

\figref{fig_pod_fc} depicts a sensing performance metric, i.e., detection probability $P_{\rm D}$, versus the power budget $P$ with a constant false-alarm probability $P_{\rm FA}=10^{-7}$.
Four targets are considered, and the probability of detection is calculated based on Eq. \eqref{eq:pod}. 
The satellite ISAC systems reduce to sensing-only ones when $\zeta$ is set to be zero, the EE performance of which is illusrated as a benchmark.
As observed from \figref{fig_pod_fc}, the ISAC systems with a larger weighting coefficient ($\zeta=0.9$) consume more power than with a smaller one ($\zeta=0.4$), to obtain a satisfying detection probability of 90\%.
By referring to Fig. 3
, it is worth noting that the ISAC systems can both enhance the communication EE and the target detection performance with a proper weighting coefficient $\zeta$.
Besides, the proposed scheme outperforms the beam squint-unaware one in the detection probability for a given power budget, as observed from \figref{fig_pod_fc}.
\section{Conclusion}\label{sec:conc}
In this paper, we investigated ISAC for massive MIMO LEO satellite systems and proposed a beam squint-aware hybrid precoding scheme exploiting sCSI, to operate communications and target sensing simultaneously.
The corresponding nonconvex optimization problem with tightly coupled variables was first converted into designing the equivalent fully digital precoder, which was handled through fractional programming.
After that, we aimed to minimize the Euclidean distance between the product of the hybrid digital/analog precoders and both the fully digital communication precoder as well as the sensing precoder.
Then, to trade off between communications and detection, a weighting coefficient was introduced and the hybrid precoders were designed through an alternating optimization framework.
Numerical results demonstrated the performance gains over the method without considering the effects of the beam squint.
Furthermore, it was shown that both the communication and sensing functions can be simultaneously supported with satisfactory performance.

\begin{appendices}
\section{Proof for the Upper bound in \eqref{eq:exrt}}\label{app:a}
For notation brevity, we introduce several auxiliary variables as $a=|\mathbf{v}_k^H[m]\mathbf{b}_k[m]|^2$, $b=\sum_{{\ell}\neq k}|\mathbf{v}_k^H[m]\mathbf{b}_{\ell}[m]|^2$, $c= N_0^{\rm c}$  and $x=\abs{g_k[m]}^2$. Then, the  logarithmic expression inside the expectation operator of the ergodic rate in \eqref{eq:exrt} can be expressed as
\begin{align}
f(x)=\log\left( \frac{ax}{bx+c}+1\right),\ x\geq 0, a,b,c\geq 0.
\end{align}
In order to verify the convexity or concavity of $f(x)$, we check its second derivative, which can be calculated as
\begin{align}
f''(x)=-\frac{ac\left((a+b)(bx+c)+((a+b)x+c)b\right)}{\left((a+b)x+c\right)^2\left(bx+c\right)^2}\leq 0.
\end{align}
This shows that the function $f(x)$ is concave with respect to the variable $x=\abs{g_k[m]}^2$ \cite{boyd2004convex}.
Then, the upper bound of the ergodic rate can be derived by utilizing Jensen's inequality, i.e., $\mathbb{E}\left\{f\left(X\right)\right\}\leq f\left(\mathbb{E}\left\{X\right\}\right)$, which is given by
    \begin{align}
    &R_k[m]=\mathbb{E}\left\{\log\left(1+\frac{|\mathbf{b}_k^H[m]\mathbf{h}_k[m]|^2}{\sum_{\ell\neq k}|\mathbf{b}_{\ell}^H[m]\mathbf{h}_k[m]|^2+N_0^{\rm c}}\right)\right\}\notag\\
    \leq& \bar{R}_k[m]\notag\\
    \triangleq&\log\left(1+\frac{|\mathbf{b}_k^H[m]\mathbf{v}_k[m]|^2\mathbb{E}\left\{|g_k[m]|^2\right\}}{\sum_{{\ell}\neq k}|\mathbf{b}_{\ell}^H[m]\mathbf{v}_k[m]|^2\mathbb{E}\left\{|g_k[m]|^2\right\}+N_0^{\rm c}}\right)\notag\\
    =&\log\left(1+\frac{\gamma_k|\mathbf{v}_k^H[m]\mathbf{b}_k[m]|^2}{\sum_{{\ell}\neq k}\gamma_k|\mathbf{v}_k^H[m]\mathbf{b}_{\ell}[m]|^2+N_0^{\rm c}}\right).
    \end{align}
\end{appendices}

\end{document}